\documentclass[fleqn,12pt]{wlscirep}
\usepackage{setspace}
\usepackage[utf8]{inputenc}
\usepackage[T1]{fontenc}
\usepackage[utf8]{inputenc}
\usepackage[english]{babel}
\usepackage{amsmath}
\usepackage{bm}
\usepackage{float}
\usepackage{cleveref}
\usepackage{lscape}
\usepackage{subfig}
\usepackage{authblk}
\usepackage[ruled,vlined]{algorithm2e}
\crefname{subsection}{subsection}{subsections}
\crefname{algocf}{alg.}{algs.}
\Crefname{algocf}{Algorithm}{Algorithms}
\usepackage{graphicx}
\title{Emergent poverty traps and inequality at multiple levels impedes social mobility}

\author[1]{Charles Dupont}
\author[1,*]{Debraj Roy}
\affil[1]{Computational Science Lab, Informatics Institute, Faculty of Science, University of Amsterdam, 1098 XH, The Netherlands}
\affil[*]{Corresponding author: Debraj Roy, Science Park 900, 1098XH, Amsterdam, The Netherlands, Email: d.roy@uva.nl}

\begin{abstract}
At the top of the Sustainable Development Goals sit two important and unresolved global challenges of our modern era: eradicating extreme poverty and reducing inequality. Although extreme poverty has decreased in the last three decades, the global poverty rate has recently increased from 7.8 to 9.1 percent due to the pandemic and climate shocks. Therefore, identifying the mechanisms at the micro (individual) and meso (community) levels, their relationship with social mobility, and implementing effective intervention strategies are critical research objectives to achieve sustainable development targets. Numerous models have been developed to examine the complex social interactions giving rise to inequality and persistent poverty, yet few approaches include multilevel dynamics. Here, we introduce a heterogeneous agent-based model to identify conditions underlying poverty traps at different levels. Three distinct regimes emerge in our model. The first regime is a single equilibrium poverty trap characterized by low wealth and an ineffective response to poverty alleviation interventions. The second regime is a double equilibrium trap that displays high wealth inequality and responds relatively well to interventions. In the third regime, all agents prosper economically. We identify key individual (behavioral) and community (institutional) mechanisms that are important for achieving sustainable reductions in poverty and inequality. At the individual level, behavioral characteristics like risk aversion, attention, and saving propensity can lead to sub-optimal diversification and low capital accumulation. At the community level, institutional drivers such as lack of financial inclusion, access to technology, and economic segregation are key drivers of inequality and poverty traps. Our results show that addressing the above factors can yield 'double dividend' -reducing poverty and inequality within-and-between communities and create a positive feedback loops that can withstand shocks. Finally, we demonstrate that our theoretical model can be used as a sandbox for cost-benefit analysis of intervention strategies.
\end{abstract}
\begin{document}
\maketitle

\section{Introduction}
At the top of the Sustainable Development Goals sits one of the most important and unresolved global challenges of our modern era: to eradicate extreme poverty and reduce inequality within and among countries. Although poverty rates have decreased over the years, nearly 10\% of the world's population still lives in extreme poverty, characterised by a lack of basic needs such as access to food, clean drinking water, education, and information \cite{globalextremepoverty}. Emerging yet inconclusive evidence suggests that COVID-19 may have put a dent in this positive trend. Estimates for 2023 indicate that poverty rates have likely returned to 2019 levels, with about 691 million people (8.6\% of the world population) living in extreme poverty. The pandemic has also caused the largest rise in between-country inequality in three decades, and reducing both within- and between-country inequality is a scientific and a policy challenge. In addition, climate change is proving to be a major obstacle for impoverished households and communities to adapt and develop in a sustainable manner. While economic inequality and poverty are distinct concepts, they are closely related. High levels of economic inequality can exacerbate poverty by concentrating wealth and opportunities among a small segment of the population, leaving others with limited resources and prospects for advancement. Poverty can act as a barrier to economic mobility, as individuals born into poverty often face systemic obstacles such as inadequate access to education, healthcare, and employment opportunities. Limited economic mobility can perpetuate cycles of poverty, where individuals and possibly entire communities remain trapped in low-income circumstances across generations. Given the distribution of wealth determines access to technology, a major technological leap could push global inequality to chasmic levels \cite{mirza2019technology}. Reflecting on the trajectory of economic development reveals a transition from a state characterized by extreme poverty to one marked by pervasive inequality. The empirical evidence highlights two key facts:

\subsection*{Multi-level poverty traps} The alleviation of extreme poverty has proceeded at a sluggish pace in Sub-Saharan Africa, and trends indicate an uptick in poverty levels within the Middle East and North Africa regions, indicating the presence of ``poverty traps'' (see Appendix B for a brief review). This trap involves self-reinforcing mechanisms that keep regions, countries, and neighbourhoods in a cycle of poverty. Barrett et al. \cite{barrett2013economics, barrett2006fractal} and Adato et al. \cite{adato2013exploring} provide evidence of multiple equilibrium in asset dynamics in rural Kenya, Madagascar, and South Africa. Evidence from rural China shows that living in poor areas diminishes the productivity of a farmer's investments, with inadequate geographical capital such as limited rural infrastructure potentially trapping households in poverty \cite{jalan2002geographic}. Previous research has observed that poverty traps often operate across multiple levels, prompting the conceptualization of multi-level or fractal poverty traps \cite{barrett2006fractal}. The implication of the poverty trap at multiple level for policy is significant, suggesting that cross-level interactions can reinforce or mitigate poverty traps \cite{poverty_traps_across_levels, economics_of_poverty_traps}. 

\subsection*{Persistent horizontal inequality} 
Inequality appears at various spatial scales, from global to local, affecting regions, countries, cities, and neighbourhoods differently. In recent decades, while global inequalities between countries have decreased, inequalities within countries have surged. For example, China's sudden shift from extreme poverty to extreme inequality is far from being an exception, as many other countries today face similar challenges \cite{china_lifting_poverty, china_income_inequality,rising_inequality_website}. Within cities, inequality can be stark between different neighbourhoods (e.g. formal vs. informal \cite{roy2018survey,mutlu2023capitalized}) and group identities (e.g., ethnicity, religion \cite{roy2018spatial}). This is particularly important because inequalities within (vertical) and between (horizontal) local communities are the ones people feel most directly. There are many persistent horizontal inequalities in developing countries: such as northern groups in Nigeria or Ghana; Somalis in Kenya; Hutus in Rwanda and Burundi; or Muslims in India. Inequality can stem from differences in education \cite{stiglitz1973education, galor1993income}, access to resources \cite{goderis2011natural} and technology \cite{aghion1999inequality}, individual risk aversion \cite{banerjee1991risk, aghion1992distribution} and saving propensity \cite{chakraborti2000statistical, banerjee2010universal}, globalization and government policies \cite{chong2007inequality}. Research on economic inequality (see Appendix A for a brief review) has expanded substantially – from Kuznet's curve \cite{kuznets2019economic}, to the Cambridge equation \cite{pasinetti1962rate} to more recently Piketty’s $r > g$. Piketty \cite{piketty2015putting} attributes the recent increase in wealth inequality to the rate of return to capital overtaking the economic growth rate.
  
While there is a clear evidence of poverty traps and persistent inequality at multiple levels, the impact of cross level interactions in poverty and inequality is not very well understood. What accounts for these differences in the dynamics of horizontal and vertical inequalities and what is the right level for intervention? A central question of interest is whether inequality is shaped by inherent mechanisms or are influenced by the interplay between intrinsic disparities and frameworks such as institutions and markets. And, if it is the latter, what is the precise relationship between these foundational factors and socioeconomic policies? Understanding these mechanism can help achieve widespread prosperity and low inequality. In response to these challenges, we have developed a two-level heterogeneous agent-based model to study the linkages among individual behaviour, community structure, market imperfections on emergent inequality and poverty at individual and community levels. Individual agents are connected in a social network from which we extract communities. We construct an economy where heterogeneous agents experience uninsurable idiosyncratic shocks and smooth consumption. Agents are risk-averse and have identical preferences deriving utility from consumption and can engage in two types of productive projects: a safe but low-return and several high-risk, high-return projects. Agents within the same community are able to form self-financing groups and undertake risk-sharing, joint investment projects. Agents have an attention mechanism that allows agents to periodically update their investment portfolios according to observed realisations of project returns (see Methods section for details).

In order to study the model's behaviour, we construct Monte Carlo (Sobol) samples and run several repetitions for each combination of parameters in order to capture stochastic effects \cite{saltelli_sampling} (see subsection Experiments in Methods for details). Then, we categorise the model runs at both the micro (agent) and meso (community) levels according to the evolution of wealth trajectories. We find robust evidence with all experimental setups for the existence of three distinct regimes: the first is a single equilibrium poverty trap characterised by relatively low wealth inequality and wealth trajectories that all gravitate towards extreme poverty; the second is a double equilibrium poverty trap characterised by greater wealth inequality and the ability of some agents or communities to escape poverty. The application of a manifold learning-based GSA method (see Appendix C) allows us to identify critical parameter regions responsible for giving rise to each regime, and we observe important differences in the sensitivity indices of these parameters at different temporal scales as well as micro (agent), meso (community) and macro (population) scales. Intervention experiments indicate that it is not possible to extricate agents from poverty when in the single equilibrium regime, while interventions of sufficient magnitude do help to alleviate poverty to some degree for the double equilibrium regime. We find key individual (behavioural) and community (institutional) mechanisms that are important to achieve sustainable reduction in poverty and inequality. At the individual-level, behavioural characteristics like risk aversion, attention and saving propensity can lead to sub-optimal diversification and low capital accumulation. At the community-level, institutional drivers such as lack of financial inclusion, access to technology and economic segregation are key drivers of inequality and poverty traps. Our results show that targeting horizontal inequality (between-community inequality) increases the rate of successful interventions through feedback. 
\section{Methods}

In this section, we describe the key components of the agent-based model as well as relevant theory.
\subsection{Graph Construction}\label{subsec:graph_construction}
Homophily, the tendency for similar agents to form social connections, has been shown to play a significant role in how social networks are organized. Indeed, most people's social networks are relatively homogeneous with respect to socioeconomic and demographic characteristics such as race and ethnicity, age, education, and household income \cite{homophily_social_networks}. This has significant implications for the kind of information that people have access to, as well as the behaviours that we tend to adopt. In a recent paper, Talaga and Nowak propose a Social Distance Attachment (SDA) model for randomly constructing social networks based on homophily \cite{sda_homophily}. Furthermore, they show that their model is able to reproduce many of the characteristic features of social networks, making it an appealing choice for our present purposes.

First, an undirected social network consisting of $N=1225$ agents is constructed, and each agent possesses some initial amount of wealth drawn from a normal distribution centered at $\mu=10$ with standard deviation $\sigma=1$. An edge is added between two agents with respective wealth levels $w_i$ and $w_j$ with probability
\begin{equation}
    p_{ij}=\frac{1}{1+[b^{-1}d(w_i,w_j)]^\alpha},
\end{equation}
where $d(w_i,w_j)=|w_i-w_j|$,  $b$ is the characteristic distance at which $p_{ij}=0.5$, and $\alpha$ is the homophily parameter \cite{sda_homophily}. In practice, we set the characteristic distance to be the average pairwise (wealth) distance between all agents divided by 15.

Due to the probabilistic nature of edge creation, this algorithm may yield multiple disconnected components. If so, we ensure connectedness as follows: 1) identify the largest connected component and 2) for every other connected component, pick a random node and insert an edge between this node and whichever node inside the largest connected component is closest in wealth. \Cref{fig:sda_label_propagation}(a) provides an example of a small SDA graph including agent wealth levels to highlight the way in which edges are formed based on the property of homophily. The reader may refer to \Cref{fig:sda_graph_distributions} of Appendix D for some key distributions (e.g. number of communities, community sizes) for the SDA graphs that were generated.

\subsection{Community Detection}\label{subsec:community_detection}
The formation of communities based on homophily is one of the most important characteristics of social networks. A community may be defined as a subset of nodes in a graph such that ``connections between those nodes are denser than connections with the rest of the network'', and community structure significantly impacts the types of interactions and dynamics that may occur in a population \cite{defining_communities}. Concretely, agents within the same community typically interact and share information more often than with agents in different communities, which has important consequences for how a system evolves over time. Extracting a network's community structure is therefore an important problem, which we tackle by 
using a semi-synchronous label propagation algorithm proposed by Cordasco and Gargano \cite{community_detection}. This approach was chosen as it relies on how information diffuses throughout a network to identify communities, is efficient, and shown to always converge to a stable labeling.

\begin{figure}[!htbp]
    \centering
    \subfloat[]{\includegraphics[scale=0.55]{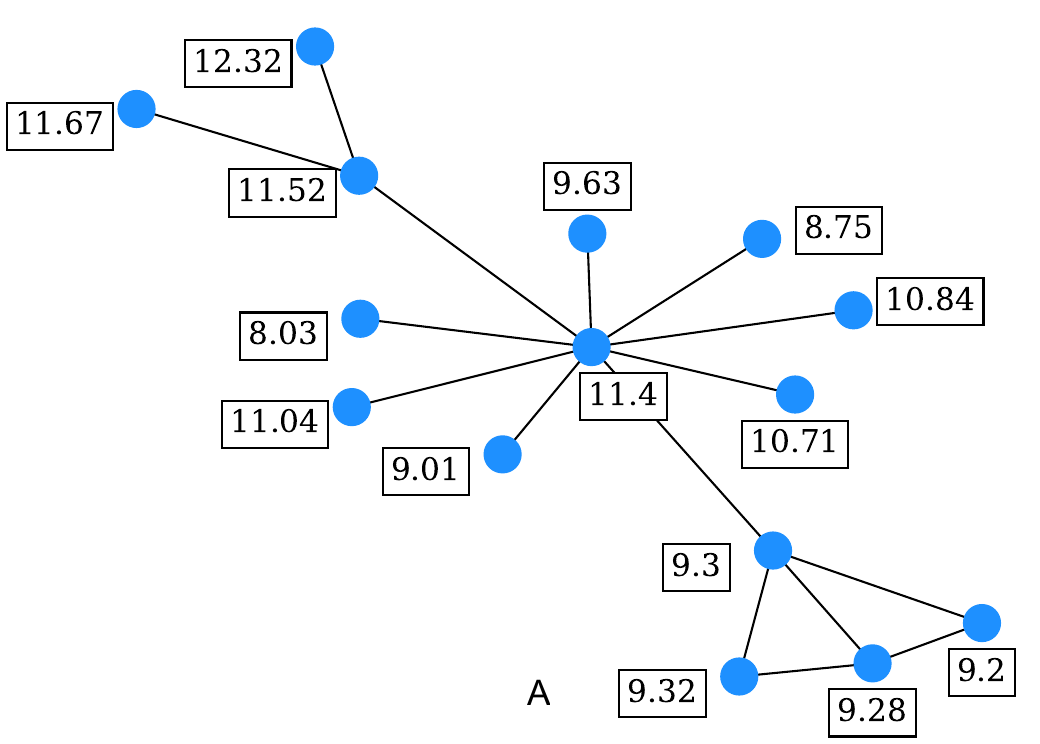}}%
    \qquad
    \subfloat[]{\includegraphics[scale=0.6]{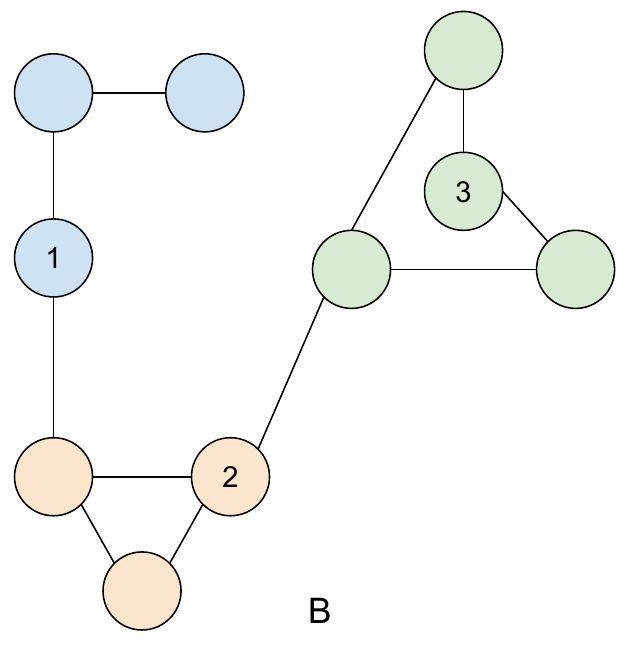}}%
    \caption{\textbf{A:} SDA graph with homophily parameter $\alpha=8$ for 15 agents with wealth values drawn from $\mathcal{N}(\mu=10,\sigma=1)$. \textbf{B:} Toy example of community detection via label propagation. Colors correspond to the result of applying the label propagation algorithm. However, as described in this section, we consider an agent to be a member of its own community as well as a member of any community that agents that it is adjacent to are part of. Hence, node 1 belongs to the blue and orange communities, node 2 belongs to the orange and green communities, and node 3 belongs only to the green community.}
    \label{fig:sda_label_propagation}%
\end{figure}

Running the algorithm after graph construction step yields a set $C$ of communities, where each element is a set of agents such that an agent cannot be in more than one community. However, we expand the notion of community membership as follows: an agent is a member of its own community as well as any community that agents it is adjacent to are part of. \Cref{fig:sda_label_propagation}(b) displays a toy example of this procedure. Node 1 belongs to the blue community and the orange community since one of its neighbours is orange. Similarly, node 2 belongs to the orange and green communities. Node 3, however, only belongs to the green community.

\subsection{Self-Financing Groups}\label{subsec:self_financing_groups}
\begin{figure}[!htbp]
    \centering
\includegraphics[width=0.7\textwidth]{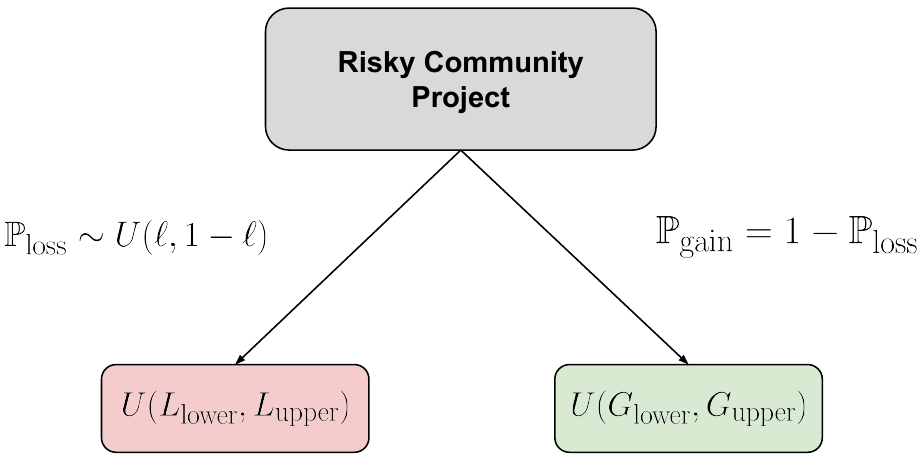}
      \caption{Diagram of a risky community project. $\mathbb{P}_{\text{loss}}$ is independently drawn once for each project, as are the random loss and gain returns in the red and green boxes, respectively.}
      \label{fig:risky_project}
\end{figure}

We allow agents of the same community to undertake joint ventures by investing in risky projects, thereby forming self-financing groups similarly to work by Gonzales Martines et al \cite{self_financing_groups}. Each community is assigned a risky project with two possible outcomes and probabilities summing to one and an agent can only invest in projects assigned to communities that it is a member of. A risky project is generated as follows. First, we randomly draw the probability that the project yields a loss from the uniform distribution $U(\ell, 1-\ell)$ with parameter $\ell\in [0.30,0.45]$. Let us call this probability $\mathbb{P}_{\text{loss}}$. The probability that the project yields a gain is then simply $\mathbb{P}_{\text{gain}} = 1-\mathbb{P}_{\text{loss}}$. Finally, we must determine the actual values of the loss and gain. The loss is drawn once per project from uniform distribution $U(L_{\text{lower}}, L_{\text{upper}})$ and the gain from uniform distribution $U(G_{\text{lower}}, G_{\text{upper}})$, with the following relations: $L_{\text{lower}} < L_{\text{upper}} < 1 < G_{\text{lower}} < G_{\text{upper}}$. \Cref{fig:risky_project} illustrates a risky project, including branch probabilities and outcomes.

In addition to community projects, each agent has the option to invest in a safe asset with a guaranteed gain of $G_\text{safe} > 1$. However, we still wish for projects to be attractive alternatives to the safe asset, so care is taken in choosing parameter values such that the expected return of each project is at least $G_\text{safe}$. 

Lastly, a project is allowed to take place if and only if the total pooled investment towards that project is at least $\theta\in(0,1)$ multiplied by the total initial wealth of agents able to contribute towards the project. If this requirement is not met, all invested funds are lost. Otherwise, the project returns the gain with probability $\mathbb{P}_{\text{gain}}$ and the loss with probability $\mathbb{P}_{\text{loss}}$.

\subsection{Portfolio Optimisation}\label{subsubsec:portfolio_optimization}
Agents must decide how they wish to apportion their capital between the safe asset and whatever risky projects are available to them \cite{de2022poverty,haushofer2014psychology}. We let the portfolio of agent $i$ be denoted $\boldsymbol{P}_i$, which is simply a vector of weights summing to one. Note that $\boldsymbol{P}_i$ is at least two-dimensional since every agent can invest in the safe asset and is a member of at least one community with a corresponding project.

In order to select a portfolio, agents must perform portfolio optimisation. This problem is approached using Cumulative Prospect Theory (CPT), which was developed in order to better explain how humans make decisions, particularly under uncertainty. The following theoretical overview of CPT closely parallels that of Luxenberg et al \cite{cptopt}. 

\begin{figure}[!htbp]
    \centering
    \includegraphics[width=0.6\textwidth]{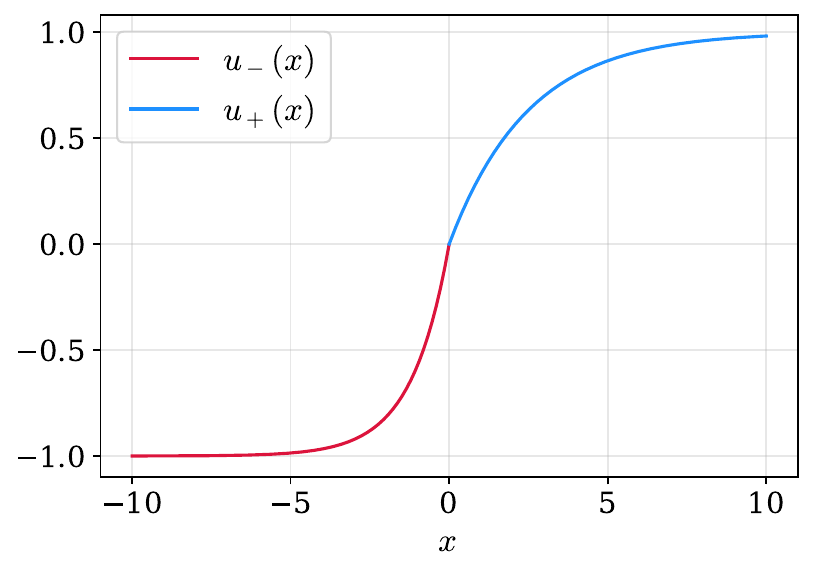}
      \caption[Visual representation of prospect theory utility function]{Prospect theory utility function with parameters $\gamma_-=0.85$ and $\gamma_+=0.4$.}
      \label{fig:cpt_utility}
\end{figure}

Agents are assumed to be risk-averse for gains. However, they are also risk-seeking for losses. For a given return $x$, the prospect theory utility function is given by
\begin{equation}\label{eqn:pt_utility_function}
    u^{\text{prosp}}(x)=
    \begin{cases}
        u_+(x) & x\geq 0\\
        u_-(x) & \text{otherwise}
    \end{cases},
\end{equation}
where $u_+(x)=1-e^{(-x\gamma_+)}$ and $u_-(x)=e^{(x\gamma_-)}-1$. There is the additional requirement that $\gamma_- > \gamma_+>0$. \Cref{fig:cpt_utility} provides a visual representation of \Cref{eqn:pt_utility_function}. The steeper exponential curve for losses (in red) can be interpreted as how a small decrease in wealth would decrease the utility more than an equal increase in wealth would increase utility. In other words, losses are felt more heavily than gains.

Next, we define some empirical distribution of returns for $n$ assets. We denote these returns $r_1, r_2,\ldots,r_N$ where $r_j\in\mathbb{R}^n$. That is, each $r_j$ is a vector of $n$ realized asset returns. Now consider some portfolio $p\in\mathbb{R}^n$, such that all weights sum to one. The resulting portfolio returns would be $r_1^\top p, r_2^\top p,\ldots,r_N^\top p$ where $r_j^\top p\in\mathbb{R}$.

Now let $N^-$ denote the number of negative portfolio returns, and let $N^+$ denote the number of non-negative portfolio returns, with $N=N^++N^-$. After sorting all of the portfolio returns, we can write the following relations:
\begin{equation*}
    p^\top\rho_1\leq\ldots\leq p^\top\rho_{N^-}<0\leq p^\top\rho_{N^-+1}\leq\ldots\leq p^\top\rho_N,
\end{equation*}
where $\rho_i$ denotes the $i^{\text{th}}$ portfolio return after sorting. In other words, $p^\top\rho_1$ is the biggest loss, and $p^\top\rho_N$ the biggest gain. Consider also the following weighing functions:
\begin{align*}
    w_+(p)&=\frac{p^{\delta_+}}{(p^{\delta_+}+(1-p)^{\delta_+})^{1/\delta_+}}\\
    w_-(p) &=\frac{p^{\delta_-}}{(p^{\delta_-}+(1-p)^{\delta_-})^{1/\delta_-}}
\end{align*}
Using these, we define the following positive and negative decision weights:

\begin{align*}
    \pi_{+,j}' &= 
    \begin{cases}
        w_+(\frac{N^+-j+1}{N}) - w_+(\frac{N^+-j}{N}) & j=1,\ldots,N^+-1\\
        w_+(\frac{1}{N}) & j=N^+,
    \end{cases},\\
    \pi_{-,j}' &= 
    \begin{cases}
        w_-(\frac{N^--j+1}{N}) - w_-(\frac{N^--j}{N}) & j=1,\ldots,N^--1\\
        w_-(\frac{1}{N}) & j=N^-
    \end{cases}.
\end{align*}

We wish for $\pi_+'$ and $\pi_-'$ to be monotonically increasing, and thereby assign greater weight to extreme outcomes. Although this is the case for most parameter choices, some can lead to the violation of this desirable property. We can ensure that $\pi_+'$ and $\pi_-'$ are non-decreasing by substituting $\pi_{+,j}'$ with $\min{(\pi_+')}$ for all $j<\text{argmin}(\pi_+')$, and substituting $\pi_{-,j}'$ with $\min{(\pi_-')}$ for all $j<\text{argmin}(\pi_-')$.

In order for $\pi_+'$ and $\pi_-'$ to each have length $N$ instead of $N^+$ and $N^-$ respectively, we zero-pad both sequences: $\pi_+=(\boldsymbol{0}^{N_-},\pi_+')$ and $\pi_-=(\boldsymbol{0}^{N_+},\pi_-')$, where $\boldsymbol{0}^k$ denotes a zero-vector of dimension $k$. 

Let us also define the following two functions operating element-wise on some vector $x$:
\begin{align*}
    \phi_+(x)&=\max{(0,x)}\\
    \phi_-(x)&=-\min{(0,x)}
\end{align*}
Using these, we can finally write the CPT utility function as follows:
\begin{equation}
    U^{\text{cpt}}(p)=f_{\pi_+}(\phi_+(u_+(\boldsymbol{R}p)))-f_{\pi_-}(\phi_-(u_-(\boldsymbol{R}p))),
\end{equation}
where $\boldsymbol{R}$ is the matrix of empirical asset returns and $f_{\pi}$ is the weighted-ordered-sum function, defined as
\begin{equation*}
    f_\pi(x)=\sum_{i=1}^{N}\pi_ix_{(i)},
\end{equation*}
where $x_{(i)}$ denotes the $i^{\text{th}}$ smallest element of vector $x$ and $\pi$ is assumed to be a monotonically increasing probability vector, which we have ensured is the case.

The CPT optimisation problem to be solved by agents is thus simply to find a portfolio $P$ maximising the CPT utility:
\begin{equation}
    \max_{\substack{P}} \,\, U^{\text{cpt}}(P),
\end{equation}
which is non-convex as shown by Luxenberg et al \cite{cptopt}. Further, Luxenberg et al \cite{cptopt} show that while CPT utility can be decomposed as a difference of two functions. The first term is a convex function with concave arguments and the second term a convex function with convex arguments. This structure allows us to derive maximum CPT utility \cite{cptopt}.

In practice, note that we assign each agent its own CPT utility function, which is parameterised as follows: $\gamma_+$ is drawn from uniform distribution $U(\gamma_+^{\text{lower}}, \gamma_+^{\text{upper}})$, $\gamma_-$ is drawn from uniform distribution $U(\gamma_-^{\text{lower}}, \gamma_-^{\text{upper}})$, $\delta_+$ is drawn from uniform distribution $U(\delta_+^{\text{lower}}, \delta_+^{\text{upper}})$, and finally $\delta_-$ is drawn from uniform distribution $U(\delta_-^{\text{lower}}, \delta_-^{\text{upper}})$.

\subsection{Portfolio Update and Attention Mechanism}
\label{subsec:portfolio_update_attention_mechanism}
An agent does not keep the same portfolio throughout the simulation. Initially, we generate $M$ random returns for each risky project as well as the safe asset (though these are constant), and each agent $i$ bases its initial portfolio $\boldsymbol{P}_{i,\text{initial}}$ on this data. However, projects yield a return at each time step, and agents can use this information to update their portfolio choices. For example, an agent may decide to reduce the amount invested towards a project that keeps failing and instead redistribute these funds towards other more promising projects or even the safe asset. 

This is done using an attention mechanism as follows. Each agent $i$ is initialised with an independent attention parameter $a_i$ randomly drawn from the uniform distribution $U(0,1)$. Then, anytime an agent updates its portfolio throughout the simulation, it computes the following weighted sum:
\begin{equation}\label{eqn:portfolio_update}
    \boldsymbol{P}_{i} = (1-a_i)\,\boldsymbol{P}_{i,\text{initial}} + a_i\,\boldsymbol{P}_{i,\text{observations}},
\end{equation}
where $\boldsymbol{P}_{i,\text{observations}}$ is the optimal portfolio of agent $i$ based on the observed project returns since the start of the simulation. Hence, the higher an agent's attention, the more importance it will place on the project returns it observes during the simulation when choosing a portfolio. 

Portfolio update times are Poisson-distributed with parameter $\lambda$, and each agent has independently sampled update times. Furthermore, we only allow for updates to be carried out after an initial period of 5 model steps.

\subsection{Consumption}
\label{subsec:consumption_choice}
At each time step $t$, each agent $i$ must consume a portion $c_{i,t}$ of its current wealth $w_{i,t}$. The amount consumed is determined by the saving propensity parameter $\beta\in(0,1)$, and is calculated as $c_{i,t}=1-\beta w_{i,t}$. Once capital has been consumed, the remainder is invested according to the agent's current portfolio, and the agent's wealth at the next time step is the return on investments (both safe and risky).

\raggedbottom

\subsection{Summary of Model Parameters}
\label{subsec:summary_model_parameters}

\Cref{tbl:model_parameters} summarises key model parameters. Parameters that we wish to vary for sensitivity analysis have their values listed in square brackets, denoting uniform intervals. Other parameters have fixed values for all simulations.

\begin{table}[!htbp]
    \centering
    \caption[Summary of agent-based model parameters]{Summary of agent-based model parameters.}
    \begin{tabular}{||c|c|c||}
        \hline
         \textbf{Parameter} & \textbf{Description} & \textbf{Value(s)}\\\hline
        $N$ & Number of agents & 1225\\
        $K$ & Number of model steps & 100\\
        $\lambda$ & Poisson parameter for generating update times & 10\\
        $\mu$ & Average of initial wealth distribution & 10\\
        $\sigma$ & Standard deviation of initial wealth distribution & 1\\
        $M$ & Initial number of project returns & 2000\\
        $L_{\text{lower}}$ & Lower uniform bound for project loss & 0.90\\
        $L_{\text{upper}}$ & Upper uniform bound for project loss & 0.95\\
        $G_{\text{lower}}$ & Lower uniform bound for project gain & 1.60\\
        $G_{\text{upper}}$ & Upper uniform bound for project gain & [1.70, 8.00]\\
        $G_{\text{safe}}$ & Safe asset gain & 1.10\\
        $\ell$ & Parameter for generating project loss probabilities & [0.30, 0.45]\\
        $\theta$ & Minimum required project investment & [0.01, 0.20]\\
        $\beta$ & Saving propensity & [0.70, 0.80]\\
        $\alpha$ & Homophily parameter & [2.00, 12.00]\\
        $b$ & Characteristic distance & see Graph Construction \\
        $\gamma_+^{\text{lower}}$ & Lower uniform bound for $\gamma_+$ & 5\\
        $\gamma_+^{\text{upper}}$ & Upper uniform bound for $\gamma_+$ & 30\\
        $\gamma_-^{\text{lower}}$ & Lower uniform bound for $\gamma_-$ & 31\\
        $\gamma_-^{\text{upper}}$ & Upper uniform bound for $\gamma_-$ & 70\\
        $\delta_+^{\text{lower}}$ & Lower uniform bound for $\delta_+$ & 0.50\\
        $\delta_+^{\text{upper}}$ & Upper uniform bound for $\delta_+$ & 0.70\\
        $\delta_-^{\text{lower}}$ & Lower uniform bound for $\delta_-$ & 0.71\\
        $\delta_-^{\text{upper}}$ & Upper uniform bound for $\delta_-$ & 0.90\\\hline
    \end{tabular}
    \label{tbl:model_parameters}
\end{table}

\Cref{alg:abm} summarises the key steps of the agent-based model. All parameters are fixed except five, which are listed at the top of the algorithm as input parameter set. After initialising global attributes such as the network, communities, projects and random initial returns, we initialise individual attributes for $N=1225$ agents. Then, we perform $K=100$ model steps during each of which agents consume some capital, invest the rest according to their portfolio, and may also update their portfolio. Before going to the next step, project returns are divided amongst investors to compute each agent's wealth for the next time step. By the end of the simulation, we have an $N\times K$ matrix of agent wealth trajectories and knowledge of agent attributes, project returns, etc.

\begin{center}
\begin{minipage}{\linewidth}
\begin{algorithm}[H]
\caption[Agent-based model for multi level poverty trap formation]{Agent-based model for multi level poverty trap formation}\label{alg:abm}
    \KwIn{parameter set $\{\ell, G_{\text{upper}}, \beta, \theta, \alpha\}$} 
    \Begin{
        Generate initial agent wealths, $w_{i,1}\sim\mathcal{N}(10, 1)$, $i=1,2,\ldots,N$\\
        Construct SDA graph using homophily parameter $\alpha$\\
        Perform community detection\\
        Determine minimum required investment cost for each community project using $\theta$\\
        Generate a risky project for each community using $\ell, G_{\text{upper}}$\\
        Generate $M$ initial project returns\\
        \For{$i=1,\ldots,N$} {
            Assign CPT Utility\\
            Assign attention parameter $a_i\sim U(0,1)$\\
            Generate set $T_i$ of portfolio update times\\
            Compute initial portfolio $\boldsymbol{P}_i = \boldsymbol{P}_{i,\text{initial}}$\\
        }
        \For{$t=1,\ldots,K$} {
            \For{$i=1,\ldots,N$} {
                \If{$t\geq 5$ and $t\in T_i$} {
                    Compute $\boldsymbol{P}_{i,\text{observations}}$\\
                    Perform portfolio update using \Cref{eqn:portfolio_update}\\
                }
                Agent chooses consumption amount $c_{i,t}$ using $\beta$\\
                Agent contributes remainder of wealth towards projects and safe asset according to their portfolio
            }
            \For{each project} {
                \If{investment $\geq$ minimum required investment} {
                    Project return = loss with probability $\ell$, or gain with probability $1-\ell$
                } \Else {
                    Project return = 0
                }
            }
            Let $\boldsymbol{R}$ be the vector of project returns concatenated with $G_{\text{safe}}$\\
            \For{$i=1,\ldots,N$} {
                $w_{i,t+1} = (w_{i,t}-c_{i,t}) \boldsymbol{P}_i\,\boldsymbol{\cdot}\boldsymbol{R}$
            }
        }
    }
\end{algorithm}
\end{minipage}
\end{center}

\subsection{Experiments}

Saltelli sampling is used to generate different combinations of parameters \cite{saltelli_sampling}. We choose to have 1024 unique values per parameter, resulting in a total of 7168 parameter combinations. Furthermore, twenty repetitions with different random seeds are carried out for each parameter combination to account for stochastic effects.

In addition to studying the resulting wealth trajectories of individuals and communities, we also examine whether a one-time capital injection can help the poorest agents escape poverty. In order to do this, we randomly select one parameter combination from the ``All Poor'' regime and one from the ``Some Rich'' regime. Then, for each one, we let our model run for 100 time steps, at which point we give 10 units of capital to the 100 poorest agents. Subsequently, we allow the model to run for an additional 100 steps and examine whether the targeted agents have managed to avoid falling into poverty. 20 repetitions are done to account for stochastic effects.

Lastly, we study the model's sensitivity to parameters using a novel global sensitivity analysis method \cite{gsa_valya,bazyleva2024trajectory} in order to compute diffusion coordinates and sensitivity indices, shown in \Cref{fig:diffusion_coordinates} and \Cref{fig:sensitivity_indices} of Appendix C respectively. The appendix also contains additional details regarding the method used.

\section{Results}
\subsection{Transition from Extreme Poverty to Extreme Inequality}

\begin{figure}[!htbp]
    \centering
    \includegraphics[width=\textwidth]{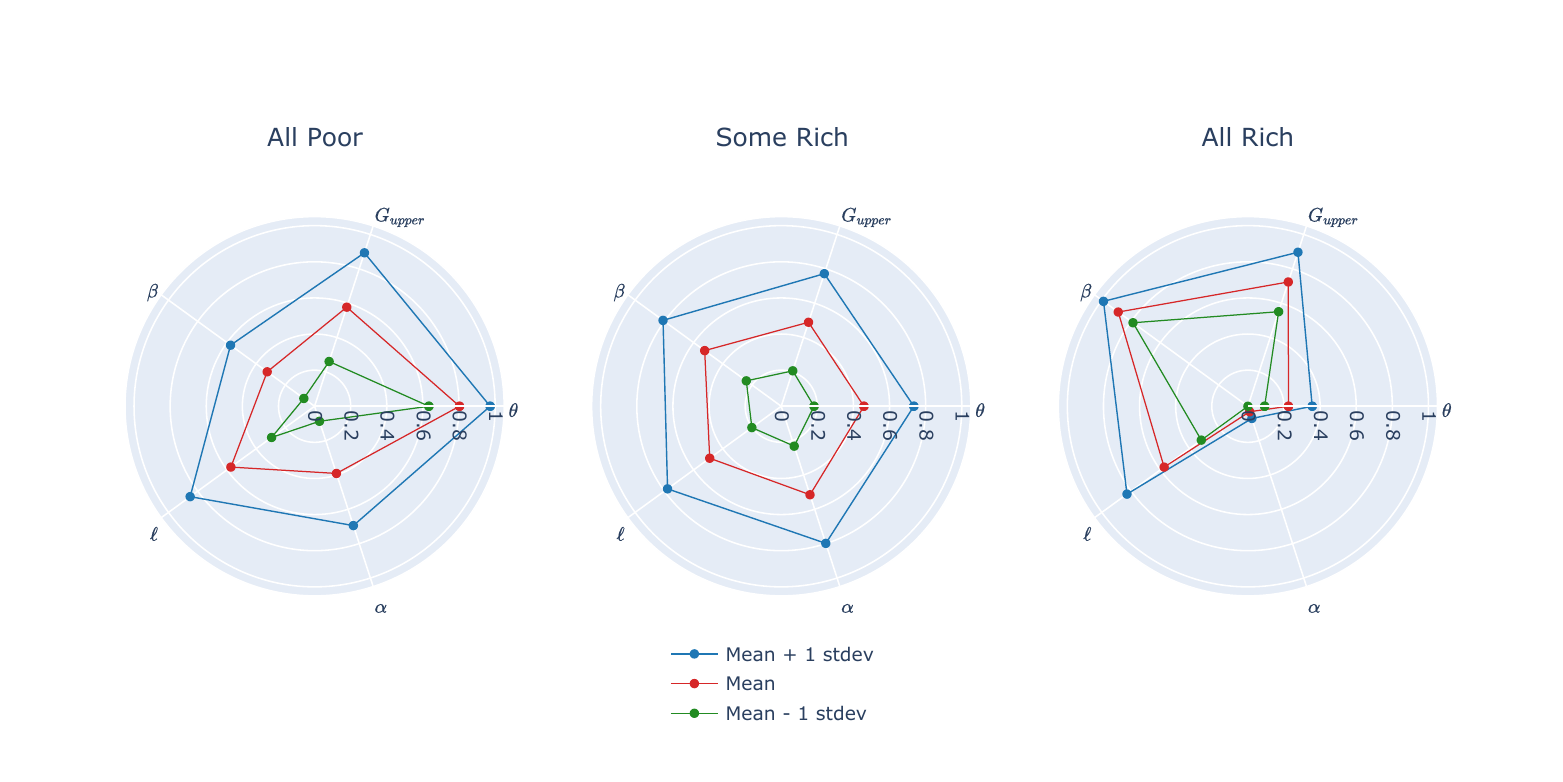}
    \caption{Average parameter values giving rise to each regime (in red). Parameter values are independently scaled between 0 (minimum value tested) and 1 (maximum value). In blue and green are 1 standard deviation above and below the average parameter values, respectively. \textbf{A:} Parameter values giving rise to the All Poor regime. High values of homophily ($\alpha$) and project cost ($\theta$) are likely to yield this regime. \textbf{B:} Parameter values giving rise to the Some Rich regime. \textbf{C:} Parameter values giving rise to the All Rich regime. Homophily and project cost are notably smaller for simulations landing in this regime.}
    \label{fig:radar_plot_individual}
\end{figure}

We identify three distinct regimes of model behaviour based on the wealth trajectories of agents across different parameter combinations and repetitions. In the first regime, which we label ``All Poor'', every agent in the population has a lower wealth by the end of the simulation compared to their starting wealth. This regime is characterised by extreme poverty and agents are unable to avoid or mitigate financial losses. In the second regime, the ``Some Rich'' regime, one or more agents have a final wealth greater than their initial wealth. Most of the time this wealth accumulation is experienced by only a small minority, thereby giving rise to significant inequality and wealth disparity. Finally, in the third ``All Rich'' regime every single agent has a final wealth greater than their initial wealth. We observe the following proportions of simulations landing in each regime: 12.24\% (All Poor), 87.7\% (Some Rich), and 0.06\% (All Rich). It is worth noting that the All Poor and Some Rich regimes correspond very closely with single and double equilibrium poverty traps as studied in depth by Barrett et. al \cite{barrett2013economics}. These findings show that the evolution of poverty and inequality (state space) is driven by various social, economic and financial structures (represented by the parameter space of the model) and can explain the observed transition from extreme poverty to pervasive inequality. \Cref{fig:radar_plot_individual} illustrates the typical parameter ranges giving rise to each of the three regimes. The influential parameters were identified using a global sensitivity analysis as described in Appendix C (see Figures \ref{fig:diffusion_coordinates} and \ref{fig:sensitivity_indices}). The scale of each parameter is adjusted using min-max normalization such that 0 corresponds to the minimum possible value of that parameter, and 1 corresponds to its maximum value (see \Cref{tbl:model_parameters} for a summary of model parameters). For each regime, the vertices along the red polygon correspond to average parameter values giving rise to that regime. Additionally, we depict values falling within one standard deviation in blue and green. The analysis of the parameter space reveals three key insights for reducing poverty and inequality.

First, parameter $\theta$, which modulates the minimum project cost required of each community in order for a project to be carried out, appears to play a significant role in shaping model behaviour. On average, $\theta$ is close to 0.8 of its maximum value in the All Poor regime (implying that project costs are high), whereas it is only 0.45 for the Some Rich regime, and 0.2 for the All Rich regime. Clearly, reducing the cost of undertaking community projects is helpful in lowering the barrier to financial well-being and tends to increase the likelihood of agents escaping poverty. The parameter $\theta$ in our theoretical model highlights the importance of financial inclusion in poverty alleviation. We show that reducing frictions related to financial inclusion can increase the level of entrepreneurship, allowing financially constrained individuals to become entrepreneurs, and move out of poverty. At the aggregate level, these effects could ultimately boost economic activity, reduce poverty, and potentially increase income equality - as observed in previous empirical studies \cite{park2015financial, omar2020does}. The parameter $\ell$ -- which determines the probability of projects yielding a loss rather than a gain -- does not appear to be different across regimes, while the upper uniform bound for the magnitude of project gains, $G_\text{upper}$, is on average significantly higher for the ``All Rich" regime. This implies that increasing project returns is more favorable to agents than increasing the frequency of project success. Second, the saving propensity parameter $\beta$ also plays an important role in giving rise to the different regimes. In the All Poor regime, $\beta$ typically takes lower values, indicating that agents consume more and invest less towards risky projects. As pointed out by Banerjee and Duflo, a ``main challenge for the poor who try to save is to find safety and a reasonable return" \cite{banerjee_lives_of_the_poor}. Indeed, in the All Poor regime, projects tend to fail more often than not, making it difficult for agents to invest their savings. However, agents in the All Rich regime tend to save more of their wealth in order to invest it towards high-return projects, which tend to be successful. This is in line with traditional perspective on micro and macroeconomic growth, which posits that increased savings, when channeled into investment, would fuel economic growth in households \cite{steinert2018saving}. Several empirical studies show that higher savings rates not only directly boost investment but also indirectly increase steady-state output, as rising savings often correlate with rising income, thus stimulating further investment \cite{karlan2014savings, do2023role}. Lastly, we find that the homophily parameter $\alpha$ needs to take on particularly low values in order for the All Rich regime to emerge. Low homophily implies that there is little economic segregation based on wealth at the start of the simulation, thereby improving the chances of poorer agents to be included in community projects with higher investments and chance of success. By contrast, in the Some Rich regime, homophily tends to be higher, which implies more segregation and leads to greater eventual inequality. The share of high-SES friends among individuals with low SES (low homophily) — known economic connectedness — is among the strongest predictors of upward income mobility \cite{chetty2022social,pena2021inequality,chantarat2012social,toth2021inequality}. A recent study identified that people endowed with high levels of economic and human capital enjoy improved accessibility and networks with a high prevalence of instrumental relations \cite{pena2021inequality}. There is essential inequality in the endowment of social capital, which augments economic inequality. Further, when inequality is socially embedded, traditional re-distributive policies demonstrate limited effectiveness \cite{pena2021inequality}.

\begin{figure}[!htbp]
    \centering
    \includegraphics[width=\textwidth]{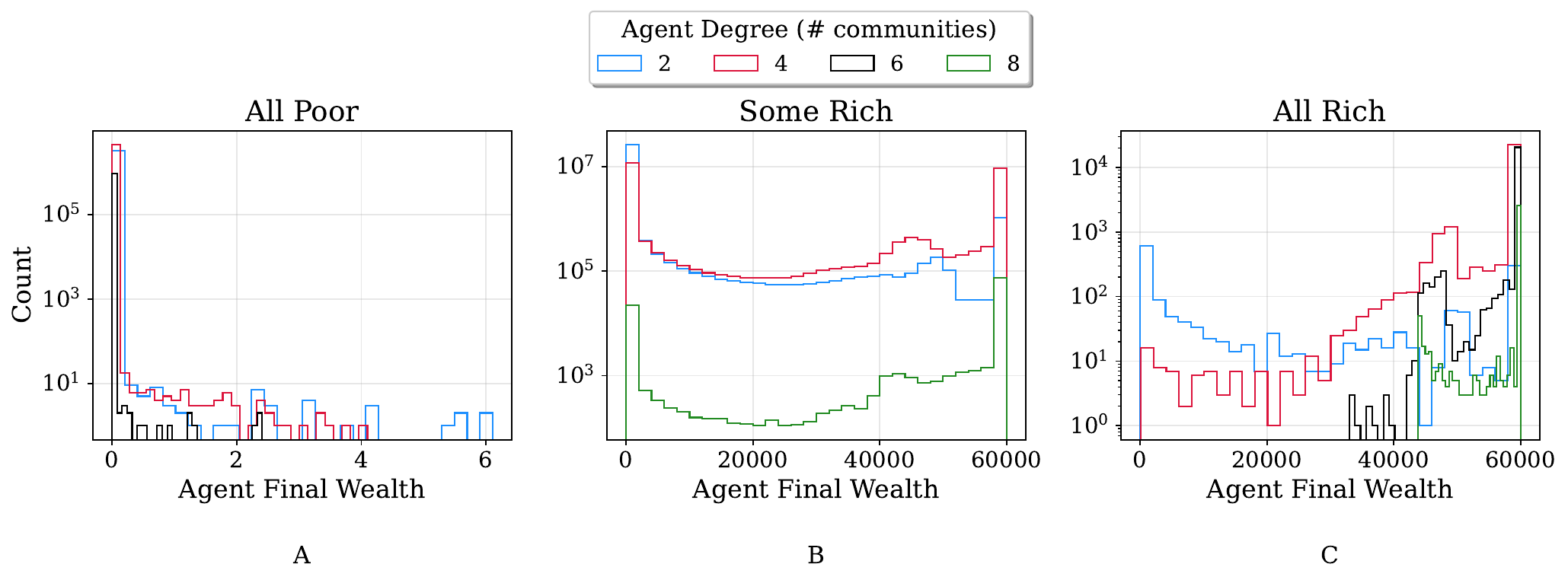}
    \caption{\textbf{A:} Distribution of final agent wealth for All Poor regime simulations. Colors correspond to the number of communities that an agent is part of. In this case, a lower degree seems to be indicative of higher final wealth. \textbf{B:} Same as sub-figure A but for the Some Rich regime. The dependence of final wealth on agent degree is less evident. \textbf{C:} Same as sub-figures A and B but for the All Rich regime. In this case a higher degree appears to be associated with higher final wealth.}
      \label{fig:agent_degree_distributions}
\end{figure}

\begin{figure}[!htbp]
    \centering
    \includegraphics[width=0.9\textwidth]{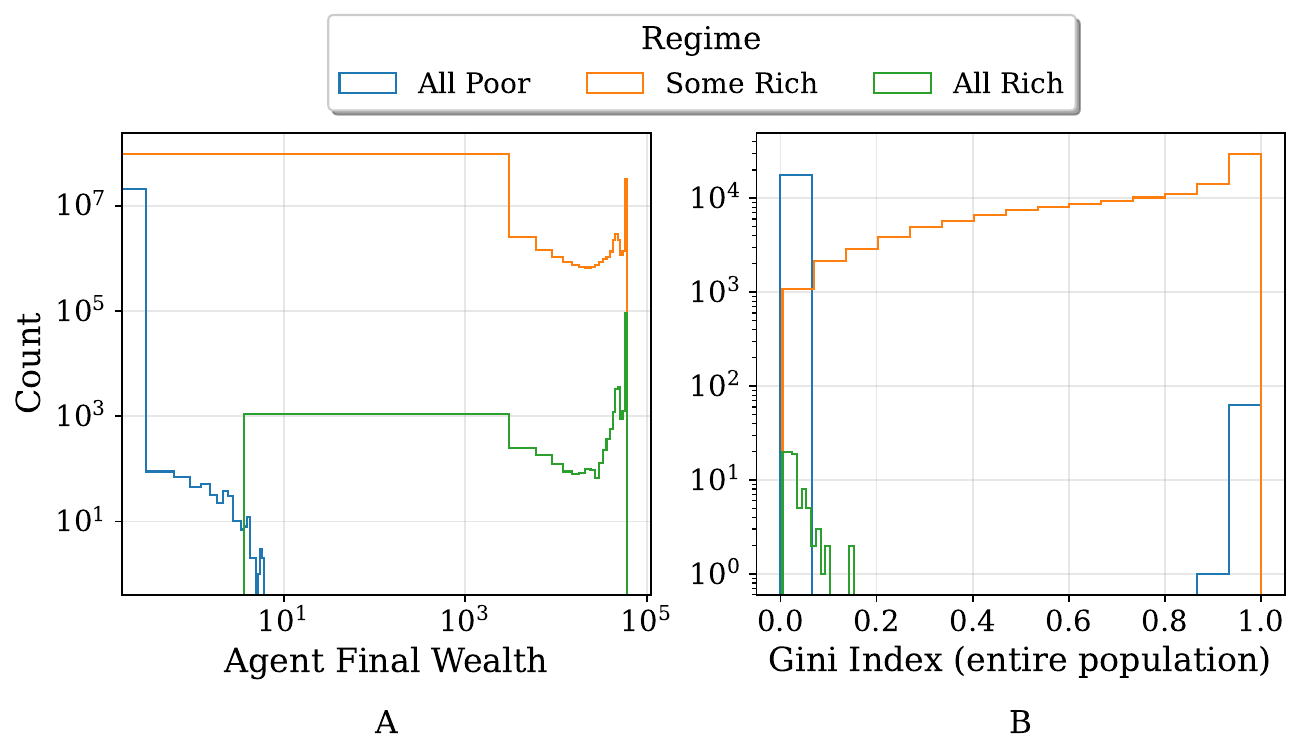}
    \caption{\textbf{A:} Comparison of final agent wealth distributions for All Poor, Some Rich, and All Rich regimes. As expected the final wealths of agents in the All Poor regime never surpass small values, whereas agents in the Some Rich regime can attain high levels of wealth. Meanwhile in the All Rich regime, final wealth levels close to zero are, by definition, impossible. \textbf{B:} Comparison of Gini indices (at final time step) for populations in different regimes. The All Poor and All Rich regimes tend to exhibit low Gini indices (signifying low inequality), which the Some Rich regime exhibits mostly high gini indices and inequality.}
      \label{fig:agent_wealth_gini}
\end{figure}

\Cref{fig:agent_degree_distributions} illustrates the relationship between the final wealth of agents in each regime and the number of projects available to the agents (the community degree). Interestingly, although all agents tend towards poverty, it seems favourable in the All Poor regime to be part of fewer communities (and thus investment projects). This can be seen from the tails of the black, red, and blue distributions extending further to the right as community degree decreases respectively. This story is completely flipped in the All Rich regime, where a higher community degree appears to lead to a higher final wealth. Given that projects on average are successful more often in the All Rich regime, it indeed makes sense for an agent to diversify their investments. By contrast, because projects fail more often than not in the All Poor regime (for instance due to higher $\theta$ values), agents tend to invest in fewer projects, thereby taking fewer risks and maintaining a larger investment towards the safe asset. This pattern of behavior is in fact supported by a growing body of literature. For instance, experimental work by Yesuf and Bluffstone with poor households in Ethiopia reveals a high degree of risk aversion and risk-averting behavior with ``perhaps significant implications for long-term poverty" \cite{poverty_risk_aversion_ethiopia}. Indeed, farmers are often prone to sub-optimal decision making due to being less willing to undertake projects with high expected returns, which results from near constant exposure to risk factors in their daily work \cite{rosenzweig_binswanger}.

\Cref{fig:agent_wealth_gini} displays the aggregated final agent wealth distributions for each regime, as well as corresponding Gini indices (calculated at the end of each simulation at the population level). While agent wealth levels quickly taper off in the All Poor distribution, both of the Some Rich and All Rich distributions have a tail of very rich agents at the far right. Moreover, the All Rich distribution does not have any data points at or near zero since agents have all attained a higher wealth level than their starting point. The Gini coefficient distributions depict a striking transition. In the All Poor and All Rich regimes, Gini coefficients are typically very low (with a few exceptions in the All Poor regime), signifying that inequality is relatively low with all agents being either poor or rich as the regime names indicate. However, the intermediate Some Rich regime is characterised by high levels of inequality, with Gini coefficients generally taking on values greater than $0.8$. Further details on the relationship between total population wealth and the Gini index for the Some Rich regime may be found in \Cref{fig:total_wealth_gini_some_rich} of Appendix F. The reader may also refer to \Cref{fig:project_returns} of Appendix E to study the distributions of project returns by regime. Overall, the transition from extreme poverty to extreme inequality found here is reminiscent of the global historical trend presented in the Introduction, wherein we argue that many nations have succeeded in reducing poverty over the last decades at the cost of increased inequality. 

\subsection{Agent Heterogeneity, Information Poverty and Sub-optimal Diversification}

In order to isolate and examine the effect of individual differences (attention and CPT utility parameters) between agents, we group agents according to the set of projects that they have access to. For instance, one group of agents could consist of those who have access to community projects 2 and 3, and another group of agents could contain those who have access to community projects 2, 3, and 5. These would be distinct groups even though all agents in the second group have access to all projects accessible to agents in the first group. 

\begin{figure}[H]
    \centering
    \includegraphics[width=\textwidth]{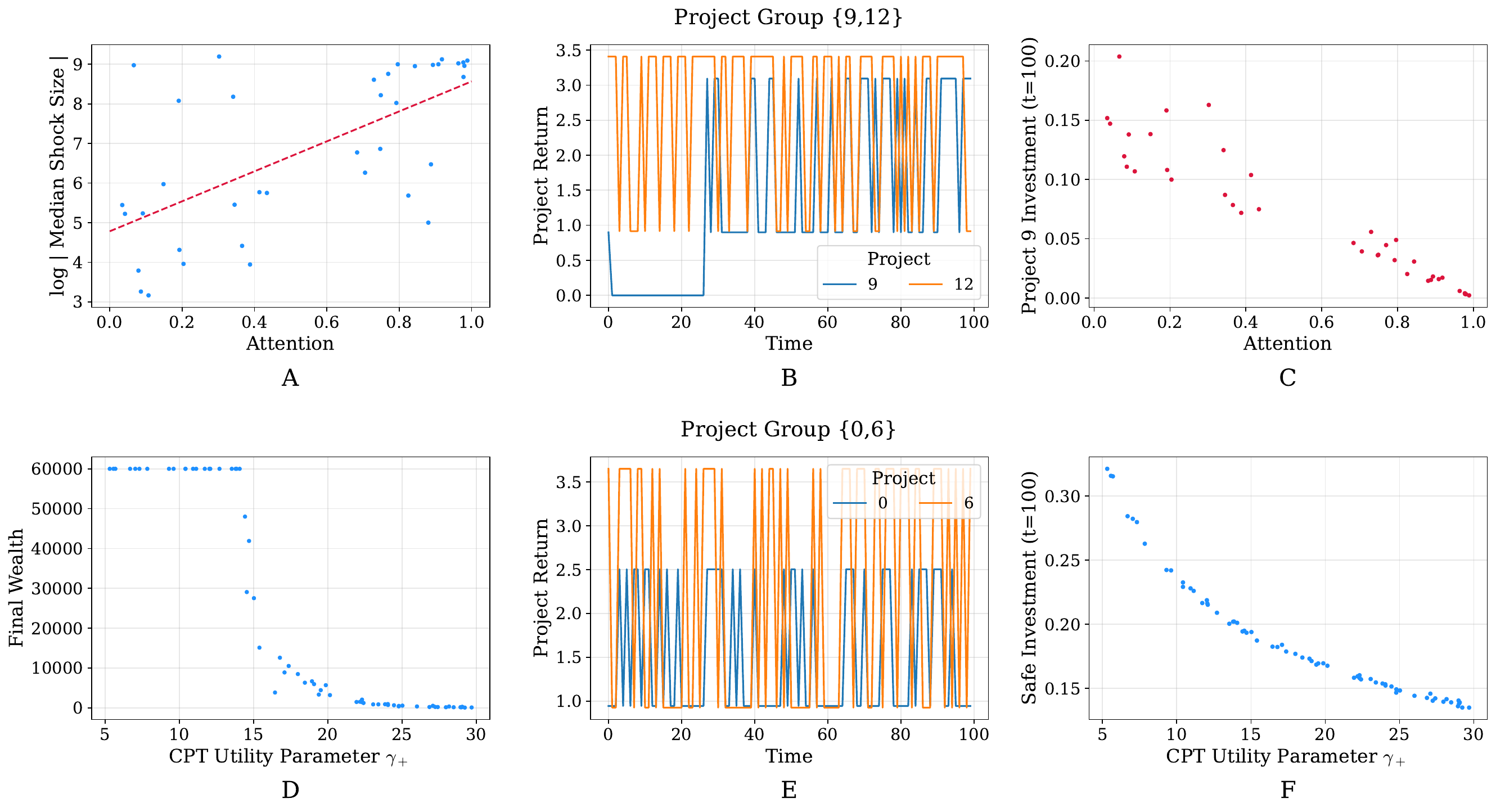}
    \caption{Sub-figures A through C all correspond to a handpicked simulation in the Some Rich regime. Sub-figures D through E correspond to a different handpicked simulation in the Some Rich regime. \textbf{A:} Impact of agent attention on the median shock size experienced for agents with access to the same two community projects (numbered 9 and 12). Agents with higher attention experienced economic shocks of greater magnitude. \textbf{B:} Returns of community projects 9 and 12. \textbf{C:} Impact of agent attention on the proportion invested in community project 9 by the end of the simulation. \textbf{D:} Impact of $\gamma_+$ CPT utility parameter on agent final wealth for agents with access to the same two community projects (number 0 and 6). \textbf{E:} Returns of community projects 0 and 6. \textbf{F:} Impact of $\gamma_+$ CPT utility parameter on the proportion invested into the safe asset by the end of the simulation.}
    \label{fig:attention_gamma_pos}
\end{figure}

Let us first pay attention to the top row of \Cref{fig:attention_gamma_pos}, where we highlight the impact of agent attention on the magnitude of economic shocks (defined as decreases in wealth between subsequent time steps) for agents with access to community projects 9 and 12 in a given simulation. The leftmost figure reveals a positive correlation between attention and the magnitude of economic shocks that agents experience -- higher attention translates to experiencing more significant shocks. In order to explain this outcome, we want to better understand how projects 9 and 12 are performing over the course of the simulation. The middle figure reveals that project 9 fails consistently for the first 25 steps or so. Agents with high attention will notice this and invest more capital towards project 12 when updating their portfolio. On the other hand, agents with lower attention will ignore these empirical returns and favor their initial portfolio. The rightmost figure demonstrates this clearly, where we notice that lower attention results in investing a larger proportion of capital towards project 9 by the end of the simulation. This investment, although initially dubious, ends up benefiting agents with low attention since they tend to have more diversified portfolio and are thus able to better absorb economic shocks compared to agents with high attention.
Now let us take a look at the second row of \Cref{fig:attention_gamma_pos} for an example of how CPT utility parameter $\gamma_+$ can impact the final wealth of agents with access to projects 0 and 6. The leftmost figure shows a steep transition where agents with $\gamma_+<15$ are extremely rich by the end of the simulation whereas agents with a value for this parameter greater than 15 converge towards a state of poverty. The central figure shows that both projects 0 and 6 yield nonzero returns throughout the entire simulation. In fact, these empirical returns agree very closely with what we would expect, unlike the previous example where project 9 failed for numerous subsequent steps. Therefore, in this scenario, attention plays a rather insignificant role when agents update their portfolios. The difference in final wealth is instead attributable to $\gamma_+$, and we can observe from the rightmost figure that a lower $\gamma_+$ value translates to a higher proportion of wealth being invested in the safe asset. In other words, for this particular group of agents, it is more beneficial to take less risk by investing in the safe asset, presumably due to the risky and volatile nature of the project returns. Although not represented in this rightmost figure, we may even expect that too little risk taking could end up being harmful (for instance if the safe investment proportion was as high as > 0.70), since agents would not end up benefiting from the high returns of the projects as much. 

To conclude this section, we remark that examples such as the ones of \Cref{fig:attention_gamma_pos} abound throughout the many simulations that were carried out. We have restricted the present analysis in order to showcase the complex and rich behaviour that can be observed using our model as a result of heterogeneity in the agent population. Importantly, whether attention and $\gamma_+$ are related to economic shocks or final wealth depends on numerous factors, such as which group of agents is considered, but also which model parameters and regimes are being examined. This outcome highlights the importance of considering traps at multiple levels, since behavior can vary so drastically between different groups or communities of agents even when they are part of the same larger population. Two key insights emerge from our analysis. First, a crucial link may also be established between the attention mechanism in our model and the notion of ``information poverty" in the literature, which is a situation wherein individuals or communities do not have the necessary skills or knowledge to obtain information, as well as interpret and apply it correctly to a particular problem \cite{information_poverty}. Agents with higher attention pay greater attention to empirical observations, implying that they have better access to information. Similarly in real world settings, individuals or communities with greater access to sources of information such as education or the internet are less likely to experience information scarcity, thereby empowering them to make more informed decisions and respond appropriately to different scenarios that may arise. While this analogy is not perfect, it does underscore the importance of what information agents have access to, and crucially, what they decide to do with it -- even minor differences in information access and information use can lead to drastically different outcomes for different agents. Second, we show that sub-optimal diversification due to information acquisition, particularly in the context of investments or economic activities, can indeed have an impact on poverty. Sub-optimal diversification can lead to higher income volatility \cite{van2010information,capuano2011causes}. For instance, if a household relies solely on one unstable source of income, such as seasonal agricultural work, they may face periods of scarcity when that income dries up \cite{michler2017specialize}. This can push them into poverty or exacerbate their existing poverty. In times of economic downturns, natural disasters, or other crises, families with sub-optimal diversification may lack resilience to bounce back, potentially slipping into poverty or facing prolonged periods of financial hardship \cite{antonelli2022crop}. At the community level a lack of diversification can can lead to widespread unemployment and poverty, if, for example a community heavily relies on a single industry that faces a downturn or technological disruption.  

\subsection{Asset-Based Interventions are Inefficient}

\begin{figure}[H]
    \centering
    \includegraphics[width=\textwidth]{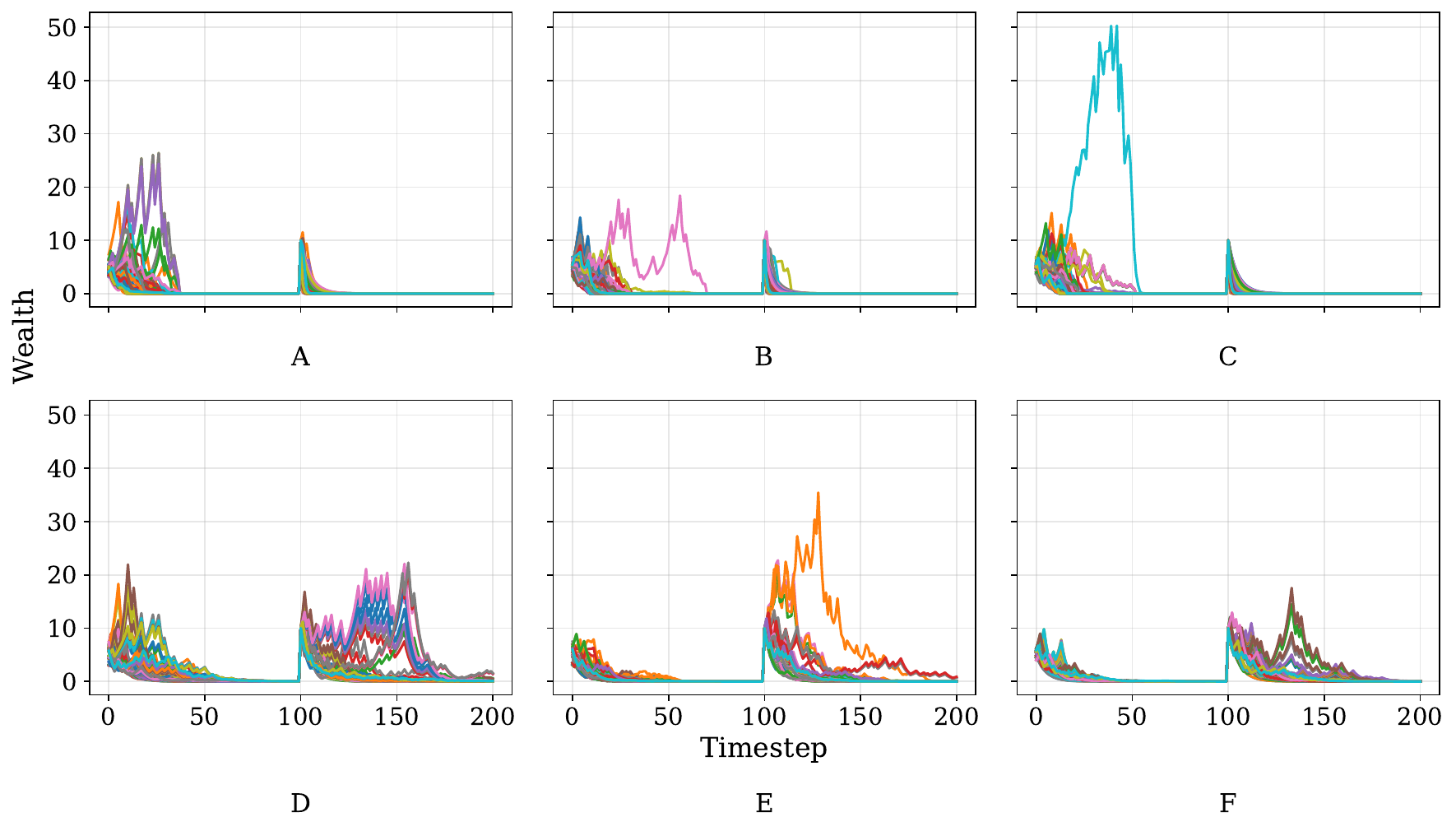}
    \caption{Examples of interventions carried out in the All Poor regime (top row) and the Some Rich regime (bottom row). In each case, 10 units of capital are given to the 100 poorest agents at the 100th time step, and the simulation then keeps running until time step 200. Interventions in the All Poor regime are found to be ineffective and targeted agents return to poverty, whereas interventions in the Some Rich regime are able to extricate some agents from poverty. This underscores the inefficacy of asset-based interventions in situations where underlying factors contributing to poverty (model parameters) are not properly addressed. \textbf{A:} First example of an intervention in the All Poor regime.  \textbf{B:} Second example of an intervention in the All Poor regime. \textbf{C:} Third example of an intervention in the All Poor regime. \textbf{D:} First example of an intervention in the Some Rich regime. \textbf{E:} Second example of an intervention in the Some Rich regime. \textbf{F:} Third example of an intervention in the Some Rich regime. }
    \label{fig:interventions}
\end{figure}

\Cref{fig:interventions} displays a sample of three out of twenty interventions that were carried out for a choice of parameters corresponding to the All Poor (top row) and Some Rich (bottom row) regimes. At $t=100$, the poorest 100 agents were identified and received a wealth injection of 10 units, after which the model was allowed to run for another 100 time steps. The trajectories displayed below are those corresponding to this subset of agents. 

For the All Poor regime, we find that in all 20 repetitions of our experiments, the targeted agents all fall back into poverty with final wealth approaching zero ($<10^{-8}$) by $t=200$. This seems to imply that interventions that do not change the underlying landscape of factors (in this case the parameter space) contributing to poverty will not favour impoverished agents in the long term. By contrast, for the Some Rich regime, we find that approximately 32\% of targeted agents manage to escape poverty by $t=200$, with a standard deviation of 13\%. Note that here we define the poverty line as being set by the wealth of the wealthiest agent among the 100 poorest agents at the end of the experiment. This outcome implies that when the underlying parameter space is conducive to upward mobility, a capital injection can prove to be somewhat successful at helping agents to escape from poverty. 

The efficacy of capital injection, or cash transfers, as a means of combating poverty remains a controversial topic even today. A recent article by Miao and Li examining the impact of cash transfers towards rural households in China casts doubt on whether this form of intervention necessarily guarantees positive long-term effects \cite{china_cash_transfers}. Although the cash transfers enabled poor rural residents to spend more on healthcare, they also led to a decrease in labor supply and no change in education spending, despite education being a significant factor in long-term poverty mitigation. Our results offer a similar point of view, especially for the All Poor regime, in the sense that cash transfers run the risk of only providing short-term relief with no concrete long-term outcomes as a result of not addressing underlying structural and systemic issues.

\subsection{Emergence of Multi-Level Poverty Traps}
Similarly to our definition of regimes for individual wealth trajectories, we may also define regimes at the community level as follows. In the ``All Poor'' regime, each community's final total wealth is less than its total initial wealth. Meanwhile, in the ``Some Rich'' regime some communities have final total wealth greater than total initial wealth and all communities have higher final total wealth in the ``All Rich'' regime. After categorizing all model runs into these community-level regimes, we construct another radar plot similar to \Cref{fig:radar_plot_individual} in order to examine the parameter ranges giving rise to each regime.

\begin{figure}[!htbp]
    \centering
    \includegraphics[width=\textwidth]{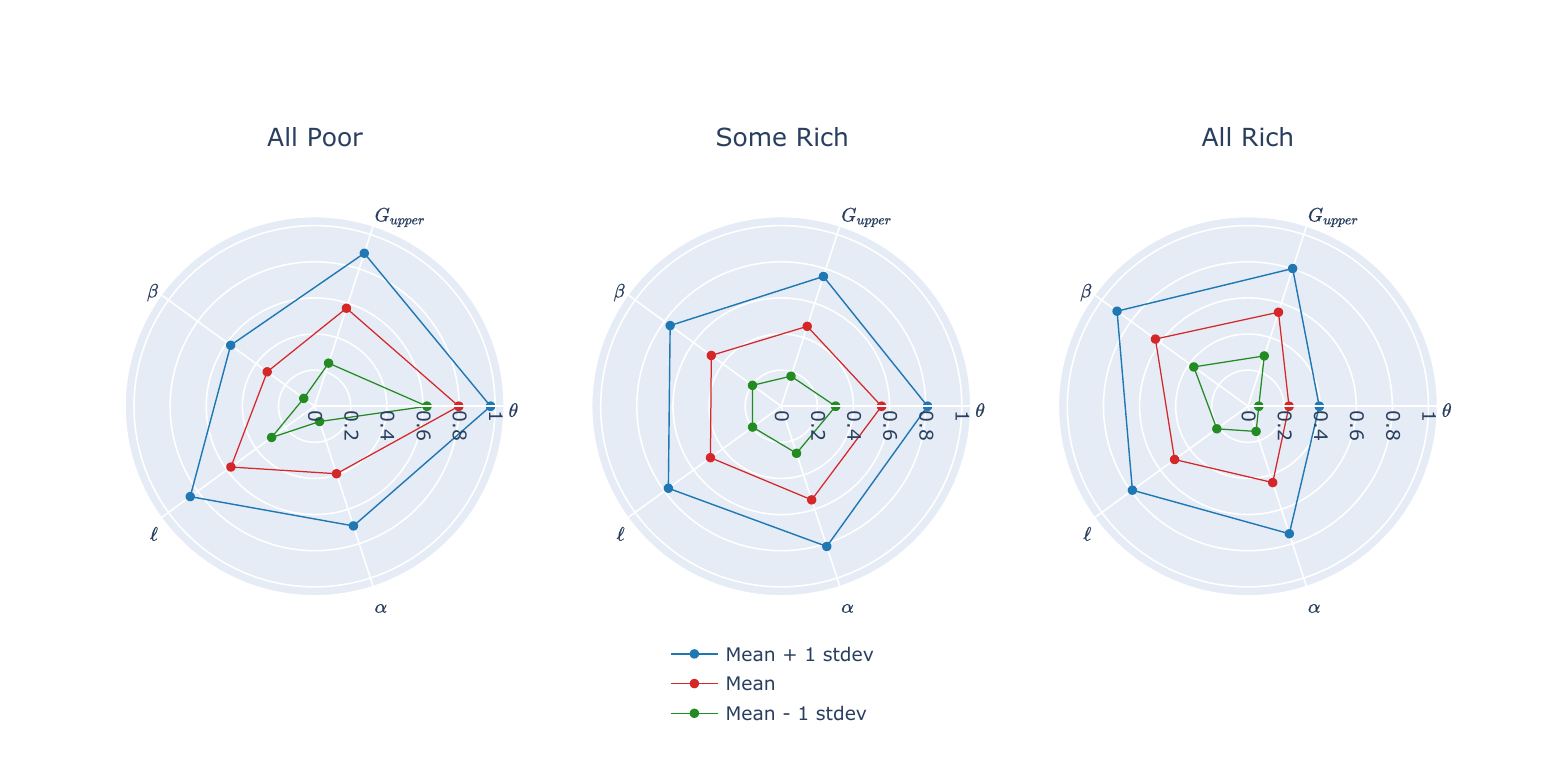}
    \caption{\textbf{A:} Average parameter values giving rise to the All Poor regime at the community level. Parameter values have all been scaled between 0 (minimum value that was tested) and 1 (maximum value). High values of project cost ($\theta$) are likely to yield this regime. Highlighted in blue and green are 1 standard deviation above and below the average parameter values, respectively. \textbf{B:} Same as sub-figure A but for the Some Rich regime. \textbf{C:} Same as sub-figures A and B, but for the All Rich regime. The effect of homophily is not as pronounced as it was for the individual-level regimes (see previous radar in \Cref{fig:radar_plot_individual}). Higher values of consumption ($\beta$) appear more likely for the All Rich regime compared to the Some Rich and All Poor regimes.}
    \label{fig:radar_plot_community}
\end{figure}

\Cref{fig:radar_plot_community} displays these results. The proportions of runs for each regime are as follows: 12\% (All Poor), 62\% (Some Rich), and 26\% (All Rich). Interestingly, we observe a much higher proportion of runs landing in the All Rich regime at the community level compared to the individual level regime definition. The All Poor and Some Rich radar plots look quite similar to those of \Cref{fig:radar_plot_individual}, but the All Rich has striking differences. Although project costs ($\theta$) still need to take on small values in order for all communities to increase their total wealth over the course of the simulation, we notice that lower homophily (small $\alpha$) is not as strong of a precondition. Furthermore, simulations giving rise to the All Rich community regime tend to have lower values of both $\beta$ (saving propensity) and $G_{\text{upper}}$ (upper bound for project returns) compared to the All Rich radar plot of \Cref{fig:radar_plot_individual}.

Interestingly, we identify $229$ simulations landing both in the All Poor community regime and the Some Rich individual regime. This indicates that it is possible for some agents to acquire significant wealth, despite all communities tending towards poverty. Conversely, we also find as many as 37,396 simulations corresponding both to the All Rich community regime and the Some Rich individual regime. This implies that we frequently find societies where all communities have a high total final wealth despite only some agents accumulating wealth. In other words, while the total wealth of communities can significantly increase over the course of a simulation, it can also be the case that many agents remain significantly impoverished. \Cref{tbl:regime_simulation_counts} summarizes the (joint) number of simulations landing in individual and community-defined regimes.

\begin{table}[!htbp]
    \centering
    \caption[Joint number of simulations landing in individual and community-defined regimes.]{Joint number of simulations landing in individual and community-defined regimes.}
    \begin{tabular}{||c|c|r||}
        \hline
         \textbf{Community Regime} & \textbf{Individual Regime} & \textbf{Count}\\\hline
         All Poor & All Poor & 17,537\\
         All Poor & Some Rich & 229\\
         All Poor & All Rich & 0\\\hline
         Some Rich & All Poor & 0\\
         Some Rich & Some Rich & 88,111\\
         Some Rich & All Rich & 0\\\hline
         All Rich & All Poor & 0\\
         All Rich & Some Rich & 37,396\\
         All Rich & All Rich & 87\\\hline
    \end{tabular}
    \label{tbl:regime_simulation_counts}
\end{table}

Overall, we conclude that our model is capable of giving rise to poverty traps at three levels. One such trap is clearly at the \textit{macroscopic scale}, where the population as a whole tends towards poverty. Interestingly, we also observe a high number of simulations where only some agents become rich and others do not, hinting at high levels of inequality on a global scale. Another set of traps is at the \textit{community level (meso scale)}. We find simulations where communities tend towards poverty, but have a few rich community members able to avoid poverty. We also have communities tending towards poverty with all community members also being poor. Finally there are mechanism of traps at the \textit{micro (individual)} level - we find simulations where communities are overall rich but some community members are stuck in poverty. In fact, as mentioned earlier, as many as 62\% of simulations give rise to societies where some communities thrive and others do not. We thus find that inequality and poverty traps can arise at the population level, but also within and between communities. This result is consistent with many empirical findings, in particular work by Barrett et al.\ where the authors argue that poverty traps can appear not only at the individual or household level, but also at community, regional, and even national scales \cite{barrett2013economics}. The multi-level poverty trap approach offers a distinct advantage by enabling an effective allocation of resources for interventions at the most appropriate levels. By understanding how interventions at different levels can have cascading effects, we can ensure targeted and coordinated efforts. Unlike relying on ad hoc decisions or anecdotal experiences, stylized modeling facilitates reasoning through various poverty alleviation strategies and outcomes, aiding practitioners in making informed decisions.

\section{Discussion}
The transition from extreme poverty to extreme inequality highlights the importance of addressing various interconnected challenges to achieve sustainable development goals (SDGs) effectively. Here, we present a pathway that outlines steps toward achieving sustainable development while addressing both poverty and inequality:

\subsection{Inequality Drives Persistent Poverty}
In the Introduction we made two observations. First, although extreme poverty has decreased in the last three decades, recently about 97 million more people are living on less than \$1.90 a day because of the pandemic, increasing the global poverty rate from 7.8 to 9.1 percent. Globally, three to four years of progress toward ending extreme poverty are estimated to have been lost. Second, between-country inequality is estimated to be increasing for the first time in a generation. Emerging evidence even shows that within countries inequality may also have worsened. The key question is – are we stuck between extreme poverty and extreme inequality? Why is progress so vulnerable to shocks?

An examination of the parameter space reveals that our model captures three extreme regimes: one characterized by poverty and the other prosperity. Both regimes have low inequality. An interesting third regime emerges where some agents are trapped in poverty while others can escape poverty and accumulate wealth. An analysis of the Gini coefficient of simulations in this regime shows that inequality is negatively correlated with economic development (see Figure \ref{fig:total_wealth_gini_some_rich} in Appendix F) – a fact well established in empirical literature \cite{oecd}. A high inequality inhibits social mobility perpetuated through increased segregation in social connections, high cost to access financial services, sub-optimal diversification and low returns to investment. Further, while there are clear pathways in the parameter space to move from the “All Poor” regime to “All Rich” via the “Some Rich” regime, our model indicates two key challenges. First, progress in poverty reduction is vulnerable to shocks due to the presence of critical transitions – where a small change in parameter space can lead to regime-shifts. Many poverty reduction programs rely on sustained economic growth and stability of the parameter space. Economic shocks, such as recessions, currency devaluations, or commodity price fluctuations, can disrupt growth trajectories and undermine the gains made in poverty alleviation. Lack of savings, access to credit, or insurance can magnify the impact of such shocks, leading to deeper and more prolonged periods of poverty. In an interconnected world, shocks in one part of the world can quickly ripple through to other regions. We show that in a multi-dimensional parameter space there are multiple pathways through which shocks can cause regime shifts in the economic landscape. Second, in unequal societies, powerful individuals or groups may engage in rent-seeking behaviour, using their influence to capture economic rents or extract resources from the rest of society without creating corresponding value. This can exacerbate inequality by diverting resources away from productive uses and concentrating wealth in the hands of a few. The intersection of inequality and political economy is a rich area of study that examines how economic policies, institutions, and power dynamics shape and perpetuate unequal distributions of resources, opportunities, and outcomes within societies. While we do not model such a feedback explicitly, the analysis of the homophily parameter in our model is an excellent proxy for the political economy and validates this finding. Finally, horizontal inequalities (see Figure \ref{fig:horizontal_inequality} in Appendix H) refer to disparities among distinct identity groups and communities, such as racial divides between blacks and whites, gender disparities between women and men, religious differences like Muslims and Hindus, or ethnic tensions like Hutus and Tutsis, among various other examples. These inequalities are inherently unjust and often endure over time. Our results indicate that Horizontal inequality is significantly higher than vertical inequality. This demonstrates that an individuals position in the distribution is mainly determined by the the community (group) they belong. Deprived groups face formidable obstacles, including limited financial resources for investing in assets and education for their children, as well as restricted access to loans. Additionally, social networks typically remain segregated within these groups, resulting in fewer beneficial connections for individuals from disadvantaged backgrounds seeking access to quality education or employment opportunities \cite{roy2018spatial}. The Sustainable Development Goals (SDGs) acknowledge the imperative of addressing inequalities not only at the individual level but also among different identity groups.

As discussed next, addressing these vulnerabilities requires a multi-dimensional approach that includes building resilience at individual, community, and institutional levels, implementing adaptive policies, and fostering sustainable development strategies that are less susceptible to external shocks.

\subsection{Pathways to Poverty Alleviation and Social Mobility}
What types of poverty alleviation interventions are possible, and how do they depend on the different pathways of inequality and poverty found at different societal scales? In recent work, Lade et al. \cite{resilience_offers_escape_from_trapped_thinking}, using a resilience-thinking lens, describe three different types of interventions that assume the existence of a barrier to escaping poverty. The first type of interventions involve capital inputs to the poor state in order to ``push it" over existing barriers, for example cash transfers or increased agricultural inputs. Although effective in some cases, it is important to note that such interventions do not lead to a corresponding systemic change and therefore may not yield substantial long-term benefits. Our intervention experiments on artificial societies in the single equilibrium (All Poor) poverty trap regime appear to replicate this particular phenomenon: mere capital injection did nothing to address the fact that undertaking projects remained too costly and that agents were consuming too much. Furthermore, as shown in our results on the impact of heterogeneity in the agent population (in the form of attention, and utility parameters), additional factors such as access to information and education can be crucially important in improving an individual's chances of escaping poverty. Further, money transfer programs, can sometimes encounter moral hazard issues. Moral hazard refers to the risk that individuals or organizations might change their behavior in response to the assurance of financial assistance, leading to unintended consequences. 

The second type of intervention consists of lowering the barrier by changing the parameters that otherwise serve as key mechanisms in inducing and reinforcing poverty. Although reducing the risk aversion of individuals and communities such that they undertake riskier ventures is technically possible in our modelling framework, in practice, this may not be the case. Individuals may not be willing to accept higher levels of risk when already facing the obstacles of extreme poverty or the near prospect of them. By contrast, we find that improving financial inclusion and access to markets or facilitating the adoption of climate-smart farming practices could realistically help lower difficulties associated with escaping the poverty cycle at the individual and community levels. Further, we show that promoting diversity can also lead to lowering the barrier. Homophily can reinforce existing inequalities by perpetuating patterns of advantage and disadvantage within social networks. For example, individuals from privileged backgrounds may have greater access to social networks with valuable connections and opportunities, while those from marginalized or disadvantaged backgrounds may face barriers to entry or exclusion from such networks. As a result, inequalities in access to resources, opportunities, and social capital can become entrenched and perpetuated across generations. Overall, diversity and access to technology empowers individuals, strengthens communities, and fosters inclusive and sustainable development. By leveraging technology effectively, policymakers, businesses, and civil society organizations can advance efforts to reduce poverty and inequality and build a more equitable and prosperous future for all.

Lastly, the third type of interventions involves major systemic transformations that bring about fundamentally different poverty and inequality landscapes. An example of such an intervention would be to provide novel developmental pathways to communities while still adhering to key tenets of social and environmental justice \cite{resilience_offers_escape_from_trapped_thinking}. One of the key findings of our model is the emergence of sub-optimal diversification. Addressing sub-optimal diversification often involves systemic transformation and broader strategies aimed at innovations (e.g. innovative agricultural practices), improving access to education and skills training (e.g. new challenge based learning), fostering entrepreneurship (e.g. micro loans, savings groups), and building resilience to external shocks. These measures can help break the cycle of poverty and create more sustainable livelihoods. Although this type of intervention is arguably the most challenging to implement, it has the potential to yield the greatest benefit since it encourages a reconfiguration of the relationships between agents, their environment, and economic practices which all contribute to persistent poverty if not appropriately addressed. For example, negative consequences could quickly arise from asset input into a community with prevalent gender inequality. While such an intervention could appear successful in terms of aggregate asset ownership, gender inequalities may intensify if women's corresponding access is not ensured in a culturally-sensitive manner \cite{UN_women_cash_transfer}.

Further, we find that the complex and multi-level nature of poverty traps makes it especially difficult to generalise findings across different contexts: a one-size-fits-all policy or intervention simply does not exist. In a recent paper, Radosavljevic et al.\ discuss the existence of fractal poverty traps wherein ``multiple low-level equilibria exist on different levels at the same time and self-reinforce through cross-level feedbacks'' \cite{poverty_traps_across_levels}. These cross-level dynamics are particularly crucial to understand because interventions at the household level, for instance, could have unforeseen and damaging consequences at the community level. This concern points to the importance of integrating information from multiple levels to better understand and quantify how multi-level interactions contribute to poverty trap mechanisms. The metric we employ to define and track progress across levels is also of critical importance. For instance, although shown to be a poor indicator of quality of life, the gross domestic product remains one of the most commonly-used metrics to compare national economies. Integrating alternative indicators such as the Multidimensional Poverty Index \cite{multidimensional_poverty_index} or the Green Growth Index, for instance, could provide a more comprehensive overview of how countries, cities, and communities are performing in terms of poverty reduction and sustainable development targets. Moreover, simply looking at the global Gini coefficient for income per capita disguises the glaring inequalities that we have argued still exist at regional and country levels. It is easy to show that a unimodal income distribution, such as the one we see today at the global level, can be constructed by summing many bimodal distributions exhibiting inequality (see Appendix G, \Cref{fig:aggregate_bimodals}).

\raggedbottom

\pagebreak
\section*{Appendix A: Inequality and Economic Modelling}
\label{app:appendix_a}

In the early 1950s, literature primarily examined the impact of growth on income distribution, neglecting reverse causality. Simon Kuznets’ seminal work addressed whether income inequality increases or decreases during economic growth. Using extensive data from 1913-1948 United States, Kuznets suggested an inverted-U relationship: inequality initially rises then declines as economic development progresses \cite{kuznets2019economic}. This Kuznets curve became fundamental in economic literature until the late 1970s, inspiring empirical validations \cite{kuznets2019economic}. Concurrently, the (post-) Keynesian school focused on functional income distribution, challenging neoclassical views. Kaldor (1955) proposed the ``Cambridge Equation,'' determining profit and wage shares based on capital accumulation and growth, diverging from marginal productivity theory \cite{pasinetti1962rate}.

Post-Keynesian contributions dwindled in significance by the late 1970s due to economic challenges like stagflation and policy failures. This led to a decline in focus on distributional concerns among economists, with attention shifting towards technical progress and growth. The rise of the rational expectations revolution, championed by Lucas, revived neoclassical principles, ignoring discussions on inequality. Endogenous growth models in the 1980s emphasized technical progress through intertemporal optimization, neglecting distributional effects. Pasinetti's observations remained relevant into the 1990s and beyond, as income distribution was often dismissed, despite its macroeconomic importance highlighted by Atkinson, Piketty, Mian, and Sufi \cite{atkinson2014handbook, piketty2015putting,mian2013household}. Most economists favoured representative-agent modelling framework that drops all the distributional considerations. Since the mid-1990s, distributional issues regained prominence in economic discourse. Atkinson and Bourguignon highlights the re-emergence of distribution as vital for development, supported by studies affirming the link between growth and inequality \cite{atkinson1987income, bourguignon2005effect}. Deininger and Squire's findings rejuvenated theoretical literature, integrating income distribution into macroeconomic and growth models by introducing rigidity to representative agent models \cite{deininger1996new}. Galor and Zeira's two-period model illustrate how initial endowments affect investment and productivity \cite{galor1993income}, while Cingano emphasizes the negative impact of inequality on human capital accumulation \cite{cingano2014trends}. Aghion and Bolton, and Piketty, explore how inequality affects credit markets and political stability, revealing its complex influence on growth dynamics and social cohesion \cite{aghion1992distribution,aghion1999inequality, piketty2015putting}. Piketty (1997) expands the Solow model by incorporating imperfect credit markets \cite{piketty1997dynamics}. This adjustment addresses the oversight of wealth distribution's insignificance in the original model. With credit rationing, interest rates and the marginal product of capital can be disrupted. Higher interest rates under these conditions lead to diminished long-term capital accumulation because credit-constrained individuals require extended periods to rebuild their capital. In the late 2000s, following the financial crisis, economists increasingly linked widening income disparities to financial instability. Fitoussi, Saraceno, Stiglitz proposed a theory based on Kaldor's framework, connecting inequality, aggregate demand, monetary policy, and financial bubbles \cite{fitoussi2010europe, stiglitz2015macroeconomic}. They argued that rising inequality was the main driver of the 2007-2008 crisis and was accompanied by structural weaknesses in aggregate demand. Income transfer from low middle-income to the wealthy, with lower consumption propensity, led to higher savings and reduced aggregate demand, potentially causing recession. Monetary expansionary policies, coupled with lax regulation, temporarily sustained consumption by lowering interest rates and increasing private debt, according to Galbraith.

Different approaches have been used in the macroeconomic literature: standard New-Keynesian Dynamic Stochastic General Equilibrium (DSGE) models; Post-Keynesian Stock-Flow Consistent (SFC) models; Macro Agent-Based (AB) Models with heterogenous interacting agents. Kumhof et al. (2015) demonstrate how changing income distributions can lead to high leverage and economic crises \cite{kumhof2015inequality}. Their DSGE model depicts financial assets backed by loans to workers, with higher inequality sparking a bubble as investors use increased income to buy more assets. This cushions workers' consumption but increases debt-to-income ratios, heightening financial fragility and precipitating a crisis. The Post-Keynesian literature has increasingly utilized Stock-Flow Consistent (SFC) models to explore the relationship between inequality, growth, and financial stability. Key contributions include Zezza (2008) \cite{zezza2008us} and Van Treeck (2013) \cite{van2014did}. Zezza's model demonstrates how growing income inequality in the US since the 1980s led to increased private expenditure relative to income, driven by rising household debt and housing market dynamics. This results in potentially unstable economic growth paths. Van Treeck (2013) extends this research by developing a three-country SFC model, calibrated for the US, Germany, and China, to analyze the impact of changing income distribution on current account imbalances. The model incorporates both personal and functional income distribution and reveals how rising household income inequality, coupled with institutional factors, explains phenomena like increased household debt and declining current accounts since the 1980s. The literature on macroeconomic agent-based models has grown rapidly, offering a complex systems approach to economic analysis with heterogeneous interacting agents. This framework proves advantageous for studying distributional matters and credit networks, considering how defaults can disrupt the entire system through feedback effects. Key works on inequality using these models include Dosi et al.\ (2013) \cite{dosi2013income}, who link income inequality to business cycle fluctuations and crises, finding that fiscal policies can mitigate these effects. Russo et al.\ (2015) \cite{russo2016increasing} show that growing inequality undermines financial stability by increasing household debt and defaults, reducing consumption smoothing and aggregate demand. 

\pagebreak 

\section*{Appendix B: Poverty Traps and Economic Modelling}
\label{app:appendix_b}
Barrett et al.\ (2013), argue that poverty traps can be generated by a variety of structural mechanisms. They could be found at the individual or household level, while some operate at the community, regional, or national scales. Some are single equilibrium mechanisms while others are multiple equilibrium mechanisms \cite{barrett2013economics}. A single equilibrium poverty trap occurs when a person or community remains poor forever, such that there is no other stable state. Whereas multiple equilibrium poverty traps exist when an individual or community fluctuates between the states of being poor or non-poor, or if there exist multiple stable states \cite{barrett2013economics}. There are one-dimensional or multi-dimensional poverty trap models depending on the number of dimensions of poverty that are being studied. Barrett et.al in their multi-dimensional poverty trap model incorporates human capital which they define as `the innate ability that includes the physical stature, cognitive development, and the level of education with which a person enter adulthood' along with financial capital. They applied the concepts of `regime shift' in two contexts namely,  `social protection' - individuals adjusting their behavior in response to poverty alleviation methods in presence of a shock, and, `social relief'- individuals doesn't change their behavior in response to the policies implemented to support individuals in the presence of a shock. Barrett et.al  gives an example of implementing `crossing barrier' behavior of regime shift in their study titled `Poverty Traps and Social Protection'. The `Need-Based Assistance' where each individual is given financial support to attain a higher income stage was implemented as a `social relief' policy and `safety nets' - used to prevent agents from falling below a threshold and `cargo net' - money transfers made to lift people above threshold, were implemented as `social protection' policy \cite{barrett2008poverty, ikegami2017poverty}. Their one of a kind model has been an inspiration for the model developed during the course of this thesis project. 

Here, we present the different types of multi-dimensional multiple equilibrium poverty trap models that exist in the literature. Among the many existing models, we can have a broad categorization as, System Dynamics Models and Agent-Based Models. Steven et.al, developed a multi-dimensional, process-based, dynamical, multiple equilibrium poverty trap model in the premise of rural agricultural communities \cite{Steven}. They considered Natural Capital, Physical Capital, and Cultural Capital as the dimensions for their multi-dimensional multiple equilibrium poverty trap model. The choice of the capitals is justified owing to the rural agricultural scenario that formed the subject of study. The models developed by Steven et.al, are unique in that they apply the resilience thinking concepts of regime shift and transformation to develop poverty alleviation methods on their multi-dimensional multiple equilibrium poverty trap model \cite{Steven}. The model formulated by Steven et.al, studied the poverty trap formation at a rural community level. They also mention how an exogenous shock like flood affects the already weak natural capital in the case of Intensification Trap Model \cite{Steven}. This in turn led to the adoption of methods that make the natural capital less susceptible to flood at the community level \cite{Steven}. Brinkmann and his colleagues developed an agent-based model to understand the dynamics of social-ecological traps in the South Western Madagascar where the farmers are negatively affected by the unsustainable exploitation of natural resources. The area witnessed a dramatic increase in the land use pressure on resources over the past years. They illustrated a model-driven scenario and devised methods to help the farmers escape this trap \cite{BRINKMANN2021103125}.

\pagebreak
\section*{Appendix C: Global Sensitivity Analysis}
\label{app:appendix_c}

In this appendix, we investigate what parameters our model is most sensitive to in order to better understand underlying model mechanics and processes. In order to accomplish this, we perform global sensitivity analysis using recent work by Bazyleva et al. \cite{gsa_valya}, where the authors employ Grassmannian diffusion maps and polynomial chaos expansion to estimate sensitivity indices. As demonstrated by the authors, this approach is particularly powerful for analysing agent-based models since such models exhibit rich temporal dynamics over multiple scales and are intrinsically stochastic.

\begin{figure}[H]
    \centering
    \includegraphics[width=\textwidth]{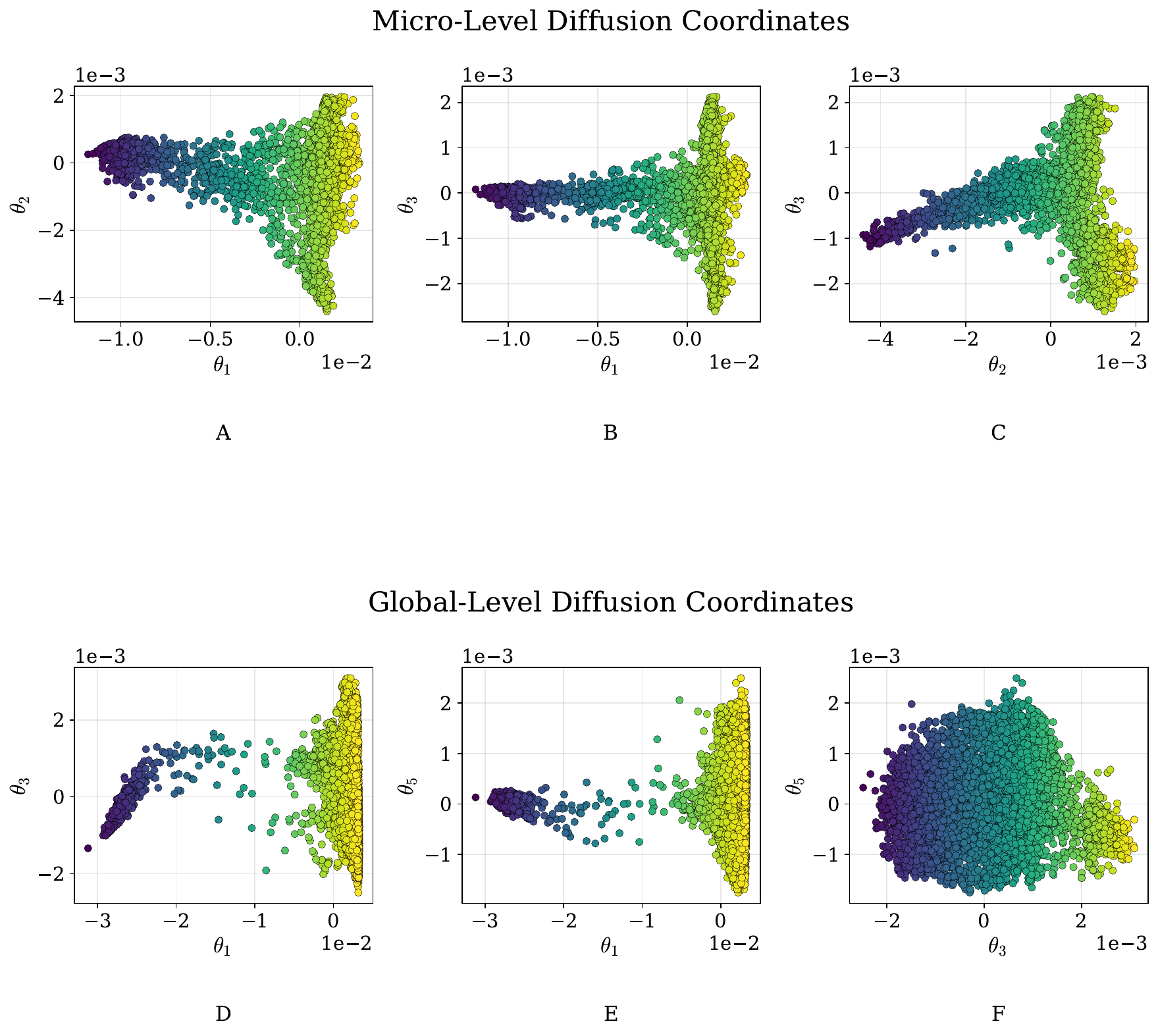}
    \caption{Diffusion coordinates obtained from applying Grassmannian diffusion maps at the micro and macro levels, similar to work by Bazyleva et al. \cite{gsa_valya}. These results are for just one of 20 stochastic repetitions of our experiment. \textbf{A:} Projection of micro-level diffusion coordinates along $\theta_1$ and $\theta_2$. \textbf{b:} Projection of micro-level diffusion coordinates along $\theta_1$ and $\theta_3$. \textbf{C:} Projection of micro-level diffusion coordinates along $\theta_2$ and $\theta_3$. \textbf{D:} Projection of macro-level diffusion coordinates along $\theta_1$ and $\theta_3$. \textbf{E:} Projection of macro-level diffusion coordinates along $\theta_1$ and $\theta_5$. \textbf{F:} Projection of macro-level diffusion coordinates along $\theta_3$ and $\theta_5$.}
    \label{fig:diffusion_coordinates}
\end{figure}

\Cref{fig:diffusion_coordinates} shows the (three-dimensional) diffusion coordinates obtained for one of twenty stochastic repetitions of our experiment. In the top and bottom rows, we find the coordinates produced at the micro, and macro levels respectively. Furthermore, the chosen manifold dimension $p$ for each level was 91 and 3 respectively.

\begin{figure}[H]
    \centering
    \includegraphics[width=\textwidth]{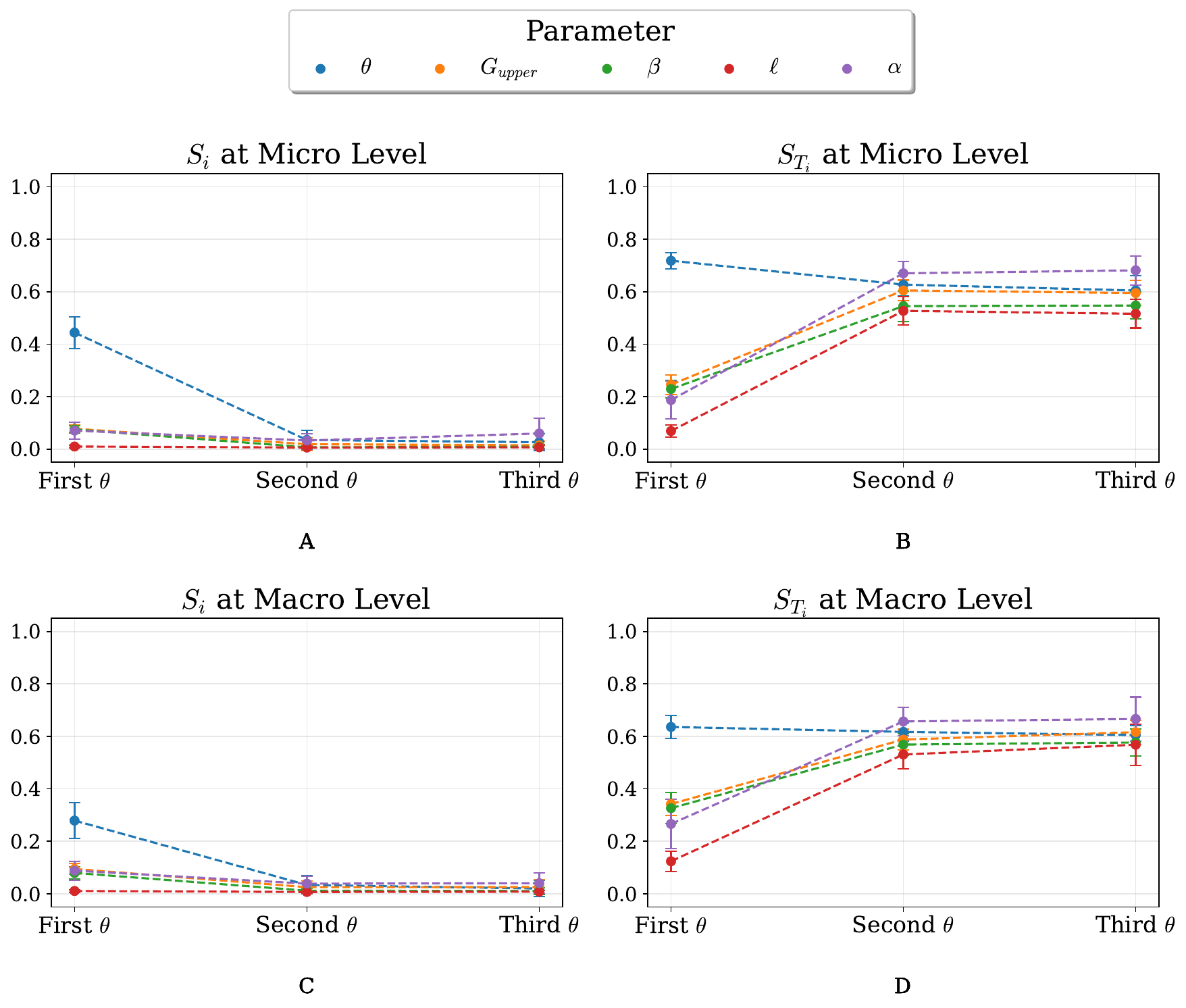}
    \caption{First and total order sensitivity indices of parameters at micro and macro levels. The maximum degree of PCE polynomials is $s=15$. The model exhibits particular sensitivity to $\theta$ (project cost) and $\alpha$ (homophily). Total order indices indicate a high degree of interaction between parameters. \textbf{A}: First order sensitivity indices of parameters at the micro level. \textbf{B}: Total order sensitivity indices of parameters at the micro level. \textbf{C}: First order sensitivity indices of parameters at the macro level. \textbf{D}: Total order sensitivity indices of parameters at the macro level.}
    \label{fig:sensitivity_indices}
\end{figure}

\Cref{fig:sensitivity_indices} displays the first- and total-order sensitivity indices of all five parameters for the top three diffusion coordinates at the micro and macro levels. It is important to point out that, in general, the first $\boldsymbol{\theta}$ corresponds to short-scale temporal dynamics, with the second $\boldsymbol{\theta}$ corresponding to longer time scales, and, finally, the third $\boldsymbol{\theta}$ mostly captures additional noise. At both the micro and macro levels, $\theta$ appears to have the highest single and total sensitivity indices along the first diffusion coordinate. This makes intuitive sense as $\theta$ modulates the investment required for a project to be carried out by a community and therefore directly impacts the system's evolution at every time step. First order sensitivity indices are near zero for all parameters along the second and third diffusion coordinates. However, all parameters have similar \textit{total} order sensitivity indices at both micro and macro levels for the second and third diffusion coordinates. In particular, we can observe that the homophily parameter $\alpha$ has highest total order sensitivity indices along these coordinates, implying that it plays an important role in longer time scales. This is also a sensible outcome, since homophily impacts network and community structure, which play a longer term role in how wealth gradually becomes distributed throughout the network. Lastly, the large differences found between first and total order sensitivity indices point to significant interactions between the parameters.

\pagebreak
\section*{Appendix D: Key Distributions for Generated Social Distance Attachment Networks}
\label{app:appendix_d}

\begin{figure}[H]
    \centering
    \includegraphics[width=\textwidth]{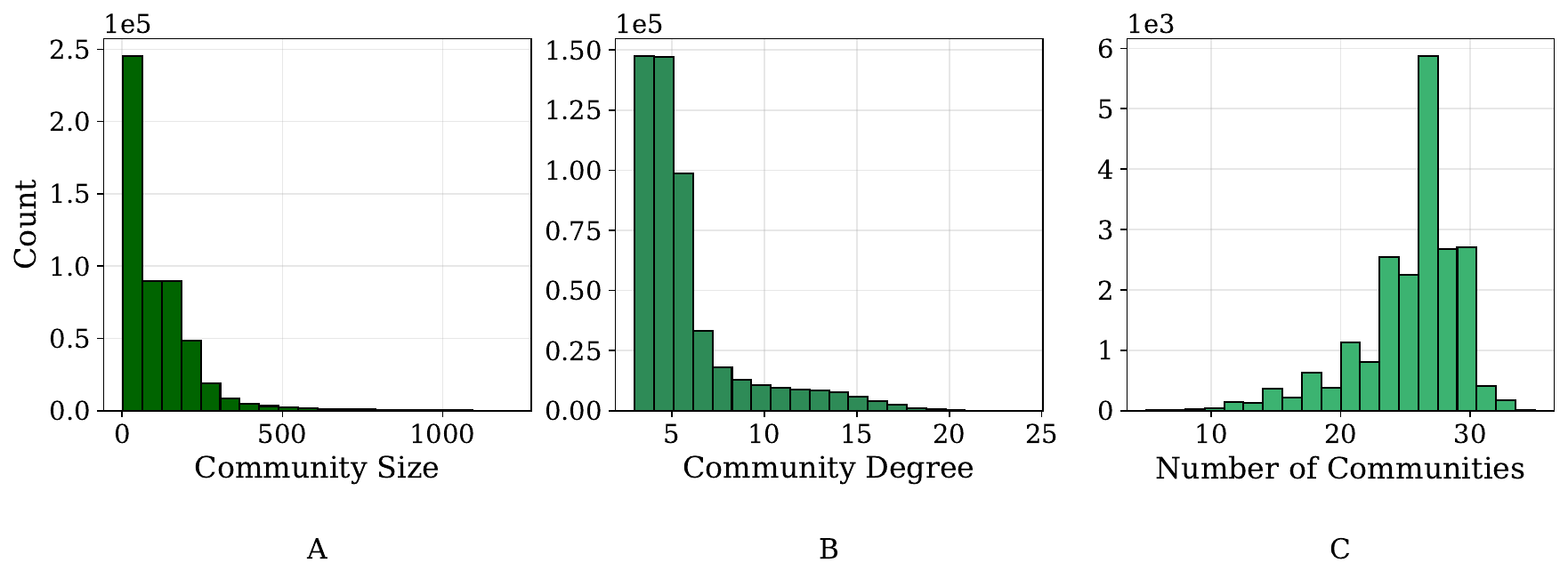}
    \caption{Distributions of interest for the social distance attachment networks generated across all experiments. \textbf{A:} Distribution of community sizes. \textbf{B:} Distribution of community degree (number of communities that a given community is connected to). \textbf{C:} Distribution of the number of communities.}
    \label{fig:sda_graph_distributions}
\end{figure}

In this appendix we analyze some important distributions for the Social Distance Attachment networks that were generated for our experiments. Looking at \Cref{fig:sda_graph_distributions} (left), we can see that communities typically have less than 250 agents, although we found an instance of a community containing 1215 agents (almost the entire population). Additionally, this distribution appears to follow a power law, which is typical of the social distance attachment mechanism. The central plot reveals that most communities are connected to only a few other communities, although some exhibit very high connectedness. Lastly, the rightmost plot shows that the number of communities in any given experiment is usually rather small, with 26 being the most common amount. Note that the community size and degree distributions follow a roughly power-law like decay.

\pagebreak
\section*{Appendix E: Distribution of Average Project Returns by Regime}
\label{app:appendix_e}

\begin{figure}[H]
    \centering
    \includegraphics[width=0.8\textwidth]{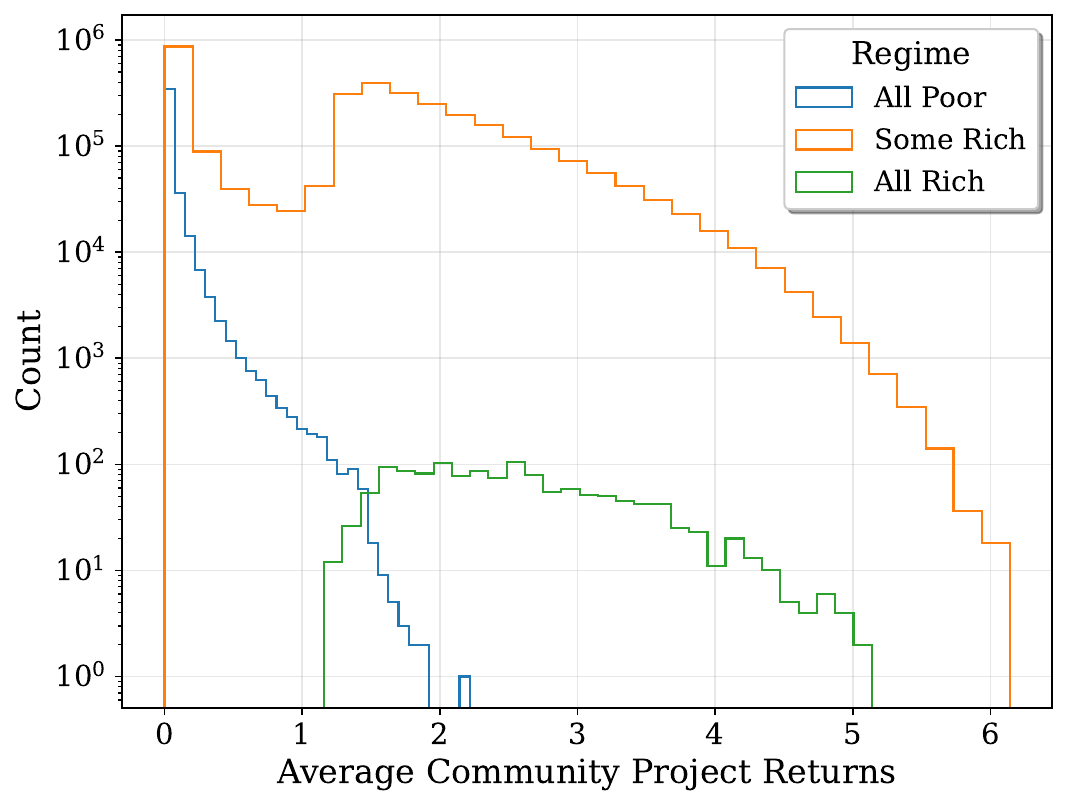}
      \caption{Distribution of average project returns across all parameter combinations and repetitions. All Rich regime projects appear to always succeed (no returns below 1), whereas the Some Rich and All Poor regimes exhibit many instances of project failure. Furthermore, while project returns decay rapidly for the All Poor regime, the Some Rich returns give rise to a more bimodal distribution with very high returns in the case of some projects.}
      \label{fig:project_returns}
\end{figure}

In this appendix we present the average project returns generated across all parameter combinations and repetitions of our experiment. This analysis occurs at the meso (community) level since projects are assigned to each community and \Cref{fig:project_returns} displays the results. In the All Poor regime, average project returns are all below 2.5, with a significant proportion being zero, which implies that projects tend to fail frequently and rarely yield gains. The Some Rich regime exhibits the largest spread of average project returns out of all three regimes. Many projects still tend to fail, but many also succeed on average and yield high gains. Interestingly, simulations in the All Rich regime do not exhibit average project returns as high as the Some Rich regime, although this could in part be due to the low proportion of simulations observed in this regime (< 0.1\%). Lastly, in this regime, project returns are always greater than 1 on average and therefore always favorable to investors.

\pagebreak
\section*{Appendix F: Relationship Between Final Total Population Wealth and Gini Index for the Some Rich Regime}
\label{app:appendix_f}

\begin{figure}[H]
    \centering
    \includegraphics[width=0.8\textwidth]{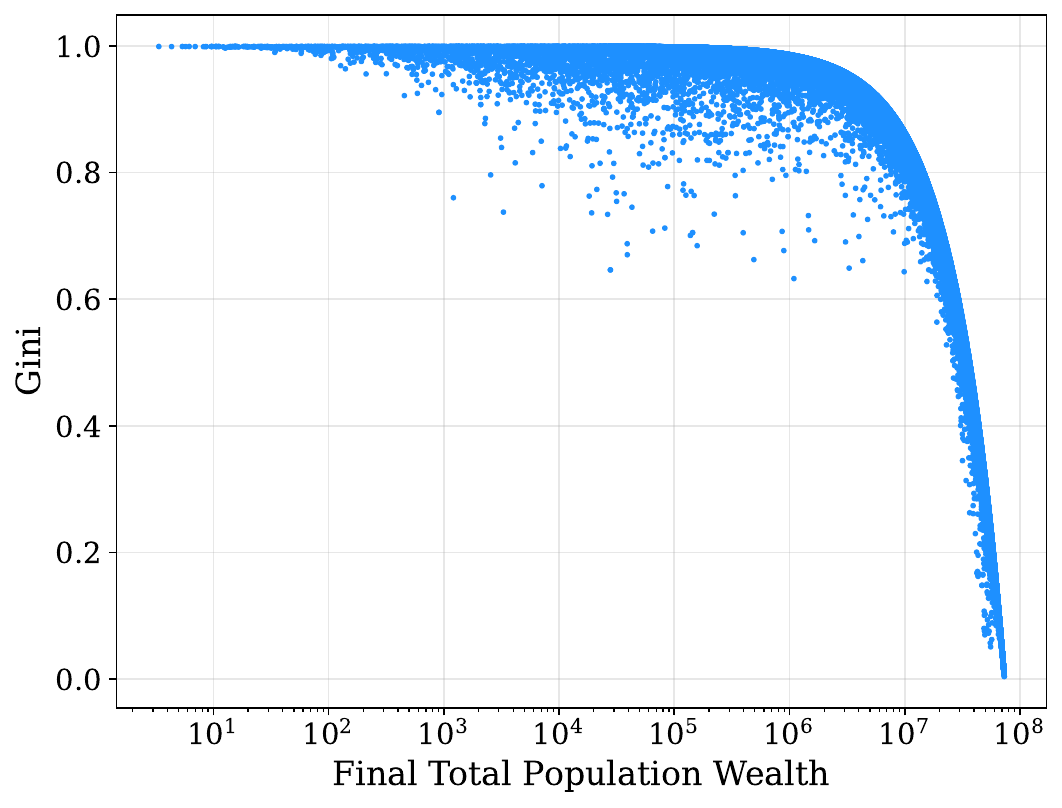}
      \caption{Relationship between total final population wealth and Gini index for the ``Some Rich" regime. As the former increases, inequality -- and therefore the Gini index -- decreases linearly (not visible here due to logarithmic scaling of the x-axis).}
    \label{fig:total_wealth_gini_some_rich}
\end{figure}

In this appendix we highlight the relationship between final wealth (aggregated over the entire population) and the Gini index for the Some Rich regime, which exhibits Gini indices across the entire spectrum of values from 0 to 1 (as seen in \Cref{fig:agent_wealth_gini}). \Cref{fig:total_wealth_gini_some_rich} demonstrates a clear decrease in Gini index as the total population wealth increases. If not for the logarithmic scaling of the x-axis, used here for improved visualization of details, we would see a very strong negative linear correlation (Pearson $r=-0.998$). This result makes sense intuitively in the Some Rich regime -- since some agents are bound to have accumulated wealth beyond their starting point, if the total final wealth is very low, then we expect a high degree of inequality, as opposed to when the total wealth is very high (meaning that almost all agents are rich).

\pagebreak
\section*{Appendix G: Summation of Bimodal Distributions}
\label{app:appendix_g}
\begin{figure}[H]
    \centering
    \includegraphics[width=\textwidth]{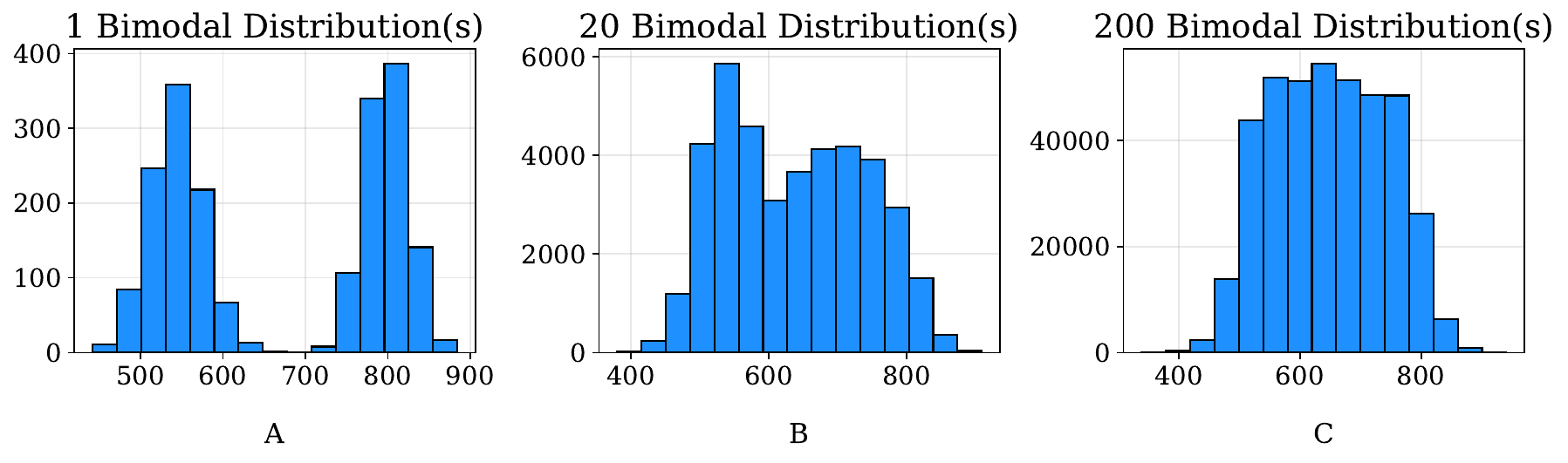}
    \caption{The sum of many bimodal distributions can appear to converge towards a unimodal one. \textbf{A:} A single instance of a bimodal distribution. \textbf{B:} Sum of 20 bimodal distributions. \textbf{C:} Sum of 200 bimodal distributions.}
    \label{fig:aggregate_bimodals}
\end{figure}

This appendix serves as a simple demonstration of how multiple bimodal distributions may be summed in order to give rise to a unimodal distribution. A bimodal distribution is generated by randomly drawing $N=1000$ samples each from two normal distributions, $\mathcal{N}_1$ and $\mathcal{N}_2$. The mean and standard deviation of these distributions are randomly drawn from uniform distributions $U_\mu(500,800)$ and $U_\sigma(10,50)$ respectively. The left plot of \Cref{fig:aggregate_bimodals} shows such a bimodal distribution. By generating and summing additional bimodal distributions, we are able to obtain a progressively tighter (middle plot) and eventually seemingly unimodal distribution (right plot). Although a toy example, this outcome is reminiscent of the way in which the global wealth and income distributions appear unimodal, thereby obscuring the high levels of within- and between-country inequality that persist even today.

\pagebreak
\section*{Appendix H: Horizontal Inequality}
\label{app:appendix_h}

\begin{figure}[H]
    \centering
    \includegraphics[width=\textwidth]{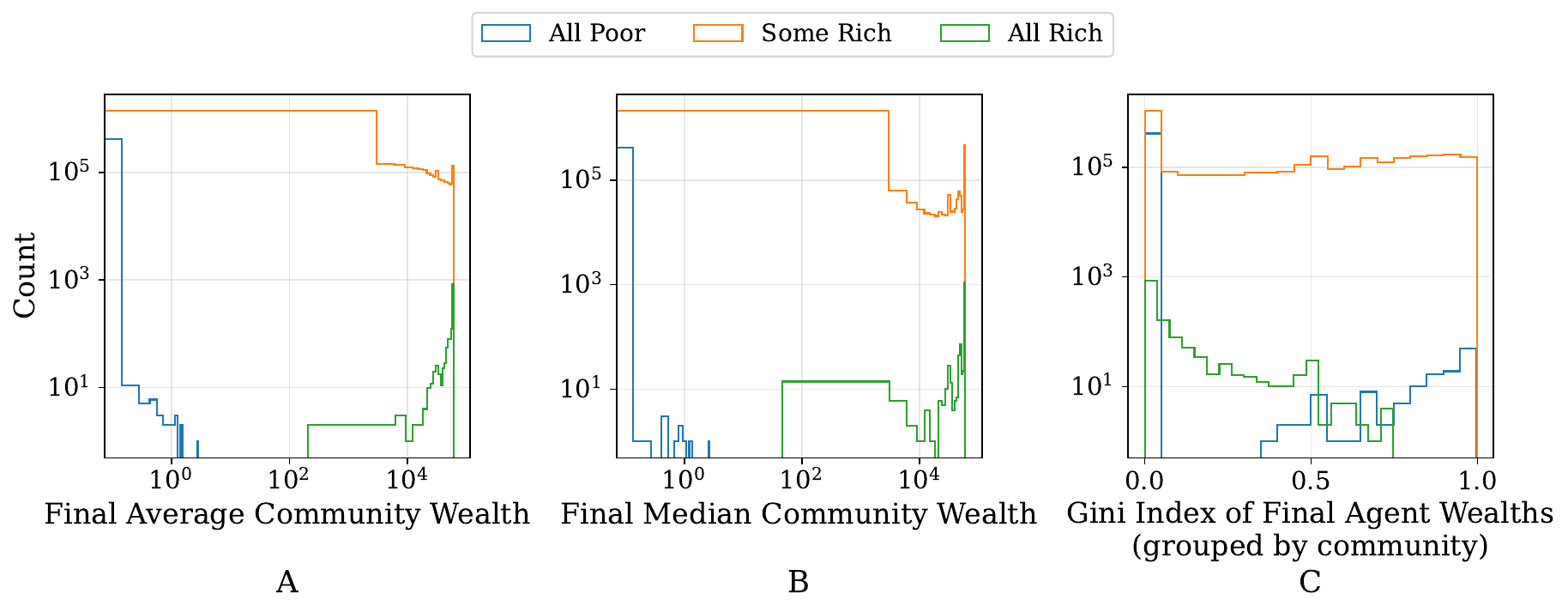}
      \caption{Similarly to our definition of regimes for individual wealth trajectories, we may also define regimes at the community level as
follows. In the “All Poor” regime, each community’s final total wealth is less than its total initial wealth. Meanwhile, in the
“Some Rich” regime some communities have final total wealth greater than total initial wealth and all communities have higher
final total wealth in the “All Rich” regime. The leftmost and central plots respectively show final average and median community wealth in these three regimes. The rightmost plot shows the distribution of inequality between communities (horizontal inequality) -- note that although the ``All Poor" regime exhibits instances of high Gini indices, this is primarily due to rounding inaccuracies (all agents have wealth close to zero, but some may have a small positive wealth of e.g. $10^{-5}$.}
\label{fig:horizontal_inequality}
\end{figure}
\pagebreak
\renewcommand{\bibname}{References}
\bibliography{sample.bib}








\end{document}